\documentclass[twocolumn,showpacs]{revtex4}%\documentstyle[prc,twocolumn,aps,epsfig]{revtex}
\usepackage{graphicx}% Include figure files
\usepackage{dcolumn}% Align table columns on decimal point
\usepackage{bm}% bold math
\usepackage{longtable}
\usepackage{epsfig}
\usepackage{amsmath}

\newcommand{\ima}{{\mbox{Im}\,}}

\begin{document}

\title{Chiral extrapolation of light resonances from 
one and two-loop unitarized
Chiral Perturbation Theory versus lattice results}

\author{J. R. Pel\'aez and G. R\'ios}

\address{Dept. F\'{\i}sica Te\'orica II. Universidad Complutense, 28040, Madrid. Spain.}

\begin{abstract}

We study the pion mass dependence of the
$\rho(770)$ and $f_0(600)$ masses and widths 
from one and two-loop unitarized 
Chiral Perturbation Theory.
We show the consistency of one-loop calculations with
lattice results for the $M_\rho$, $f_\pi$
and the isospin 2 scattering length $a_{20}$. 
Then,  
we develop and apply the modified Inverse Amplitude Method formalism
for two-loop ChPT. 
In contrast to the $f_0(600)$, the $\rho(770)$ is rather sensitive 
to the two-loop ChPT parameters -- 
our main source of systematic uncertainty.
We thus provide two-loop unitarized fits constrained by 
lattice information on $M_\rho$, $f_\pi$,
by the $q\bar q$ leading $1/N_c$ behavior  of the $\rho$ and by
existing estimates of low energy constants. These fits yield
relatively stable predictions up to $m_\pi\simeq 300-350\,$ MeV for the
$\rho$ coupling and width as well as for all the
$f_0(600)$ parameters. We confirm, to two-loops,
 the weak $m_\pi$ dependence of the
$\rho$ coupling and the KSRF relation, and
the existence of two virtual $f_0(600)$ poles 
for sufficiently high $m_\pi$. At two loops one of these poles
becomes a bound state when  $m_\pi$ is somewhat larger than $300\,$MeV. 
\end{abstract}

\maketitle

\section{Introduction.}

The spectrum of the lightest meson resonances in QCD,
particularly the scalars, is still not well understood from
first principles. Lattice QCD can provide, in principle, 
a rigorous way to extract non-perturbative quantities from QCD,
but current calculations are hindered by difficulties, like those associated 
to disconnected graphs, and their need to use relatively high quark masses, so that
appropriate extrapolation formulas are needed. In Ref.\cite{chiralexIAM}
the quark mass dependence of the mass and
width of the $\rho$ and $\sigma$ (or $f_0(600)$) mesons was predicted.
Both the mass and width were obtained from 
the poles in $\pi\pi$ scattering amplitudes generated
with the one-loop Inverse Amplitude Method (IAM) 
\cite{Truong:1988zp,Dobado:1996ps,GomezNicola:2007qj}. The IAM is obtained
from a dispersion relation based on
analyticity, elastic unitarity and 
Chiral Perturbation Theory (ChPT) \cite{Gasser:1983yg}.   The relation between $m_\pi$ and the quark mass is
well known and model independent within ChPT, which also provides
the correct $m_\pi$ dependence of low energy scattering amplitudes up to a 
given chiral order.
There were, however, two aspects
in which the calculation in \cite{chiralexIAM} needed further improvement:
on the one hand, at that time we were only able to compare
with a single lattice $\rho(770)$ mass calculation at one value of $m_\pi$. 
On the other hand, we presented a calculation
up to one-loop, namely next to leading order (NLO) unitarized ChPT,
so that the main source of systematic uncertainty, coming from higher orders, could not be 
estimated.  

In this paper we address these two issues. First,
in the next subsection we briefly review the notation and
the one-loop unitarization formalism used in \cite{chiralexIAM}. 
Next, in Sect.\ref{sec:results}
we show the good agreement of the one-loop previous calculation with
several lattice group results,
not only for the $\rho(770)$ mass at many values of the pion mass, 
but also for $f_\pi$ and the isospin-2 scattering length pion mass
dependence. In addition, since ref.\cite{chiralexIAM} 
provided a strong indication
that the $\rho\pi\pi$ coupling was almost quark mass independent, 
we perform here an explicit calculation of the $\rho$ and $\sigma$ couplings
from the residues of their associated poles, 
confirming the very weak dependence
of the $\rho$ pole on the pion mass, 
in sharp contrast with the $\sigma$ behavior.
Next, in Sect.\ref{sec:op6analysis}, we present
our calculations for unitarized two-loop Chiral Perturbation Theory.
Unfortunately, the dispersive formalism used in \cite{chiralexIAM},
which is a modification of the well known 
IAM  \cite{Truong:1988zp,Dobado:1996ps}
 that incorporates correctly the 
Adler zeros,
was developed in \cite{GomezNicola:2007qj} 
in detail only for the one-loop case.
Thus, in subsection \ref{sec:op6mIAM} we provide the
justification and the expression that modifies the IAM
to incorporate the Adler zeros correctly up to two loops.
Next, in subsection \ref{sec:op6LECS} we 
discuss the large uncertainties in the 
two-loop low energy constants, that will dominate our systematic errors,
particularly for the $\rho(770)$, as explained in subsection \ref{Ressens}.
Therefore, to ensure the correct mass dependence, in subsection 
\ref{sec:op6fits} we fit  these constants
not only to experimental data but also to the existing lattice results
with the additional constraint of respecting the $q\bar q$ leading
order $1/N_c$ behavior for the $\rho$ as well as existing estimates of the 
low energy constants. We show that all
these constraints can be accommodated with fairly reasonable low
energy parameters. Finally, in subsection \ref{sec:predictions} 
we provide predictions for the pion
mass dependence of the controversial $f_0(600)$ 
scalar resonance parameters, which are remarkably
 stable under the two-loop uncertainties.
In addition, we provide predictions for the 
coupling constant of the $\rho$ and the Kawarabayashi-Suzuki-Riazuddin-Fayyazuddin
(KSRF) relation.

\subsection{IAM and mIAM}
\label{sec:IAM}

The $\rho$ and $\sigma$ resonances appear as poles in $\pi\pi$
scattering partial waves with definite isospin $I$ 
and angular momentum $J$, in the $I=1$, $J=1$ and $I=0$, 
$J=0$ channels respectively. Elastic unitarity implies for these
partial waves and physical values of $s$:
\begin{equation}
  \label{elasticunit}
  \ima t(s)=\sigma (s) |t(s)|^2 \;\Rightarrow\; 
  \ima 1/t(s)=-\sigma (s),
\end{equation}
where $s$ is the Mandelstam variable and $\sigma (s)=2p/\sqrt{s}$, 
$p$ being the center of mass momentum. Consequently, the imaginary
part of the inverse amplitude is known exactly. ChPT amplitudes,
being an expansion $t=t_2+t_4+\cdots$, with $t_k=O(p^k)$, satisfy
unitarity only perturbatively:
\begin{equation}
  \label{unitpertu}
  \ima t_2=0,\quad\ima t_4=\sigma |t_2|^2,\quad\cdots.
\end{equation}
Let us recall that $t_2$ corresponds to a tree level calculation
with the leading order (LO) chiral Lagrangian, whereas $t_4$ contains the one loop
diagrams with LO vertices, plus tree level terms from the NLO chiral Lagrangian. The 
LO Lagrangian has no free parameters, but just 
$m_\pi$ and $f_\pi$. Higher order Lagrangians
contain Low Energy Constants (LECs) that renormalize the loop divergences and 
whose values contain the information about the underlying theory, QCD.
These LECs carry a scale dependence to cancel that of the loop integrals,
so that observables are scale independent and finite order by order.

The IAM \cite{GomezNicola:2007qj} uses
elastic unitarity and the ChPT expansion to
evaluate a once subtracted dispersion relation
for the inverse amplitude. The analytic structure of
$1/t$ consists on a right cut (RC)  from threshold to
$\infty$, a left cut (LC) from $-\infty$ to 0, and possible poles
coming from zeros of $t$. The scalar waves vanish at the so called
Adler zero, $s_A$, that lies on the real axis below threshold. Its 
position can be approximated with ChPT, $s_A=s_2+s_4+\cdots$,
where $t_2$ vanishes at $s_2$, $t_2+t_4$ vanishes at $s_2+s_4$, and so on.
We can write then a once subtracted dispersion relation for
$1/t$, the subtraction point being $s_A$, 
\begin{equation}
  \begin{aligned}
    \label{disp1/t}
    \frac1{t(s)}&=
    \frac{s-s_A}{\pi}\int_{RC}ds'\frac{\ima 1/t(s')}{(s'-s_A)(s'-s)}\\
    &+LC(1/t)+PC(1/t),
  \end{aligned}
\end{equation}
where $LC(1/t)$ stands for a similar integral over the left cut
and $PC(1/t)$ is the contribution of the pole at the Adler zero.
Note that, as $1/t$ already has a pole at $s_A$
the usual subtraction constant terms are 
actually part of the pole contribution term.

On the right cut we can evaluate exactly
$\ima 1/t=-\sigma=-\ima t_4/t_2$, as can be read from eqs. \eqref{elasticunit}
and \eqref{unitpertu}. Since the left cut is weighted at
low energies we can use ChPT to approximate $LC(1/t)\simeq LC(-t_4/t_2^2)$.
The pole contribution $PC(1/t)$ can be safely calculated with ChPT since
it involves derivatives of $t$ evaluated at $s_A$, which is a low energy
point where ChPT is perfectly justified. Altogether, we arrive to a
modified one-loop IAM (mIAM) formula \cite{GomezNicola:2007qj}:
\begin{equation}
  \begin{aligned}
    \label{mIAM}
    t^{mIAM}&=\frac{t_2^2}{t_2-t_4+A^{mIAM}},\\
    A^{mIAM}&=t_4(s_2)-\frac{(s_2-s_A)(s-s_2)[t'_2(s_2)-t'_4(s_2)]}{s-s_A},
  \end{aligned}
\end{equation}
where the prime denotes a derivative with respect to $s$ and
where we use for $s_A$ in the numerical calculations its NLO 
approximation $s_2+s_4$.
The standard IAM formula is recovered for $A^{mIAM}=0$, which is indeed the
case for all partial waves except the scalar ones. 
In the original IAM
derivation \cite{Truong:1988zp,Dobado:1996ps} $A^{mIAM}$ was neglected
since it formally yields a higher order contribution and is numerically
very small except near the Adler zero. However, if $A^{mIAM}$ is
neglected, the IAM Adler zero occurs at $s_2$, correctly only to LO,
is a double zero instead of a simple one, and a spurious pole appears
close to the Adler zero. All of these caveats disappear with the mIAM, and
the differences between the IAM and the mIAM in the physical and resonance
region are less than 1\%. 

It is important to remark that, in the above derivation, ChPT has {\it not} been used {\it at all}
for calculations of $t(s)$ for positive energies above threshold.
Note that the use of ChPT is well justified to calculate $s_A$ and $PC$,
since these are low energy calculations. ChPT has also been used to calculate
the left cut integral, which, despite extending to infinity, 
is heavily weighted at low energies, 
which once again justifies the use of ChPT.
The approximation of the left cut and the subtraction 
constants up to a given order in ChPT
 -- with the subtraction point chosen at low energy --,
together with the elastic approximation
are the only approximations used to derive the IAM from the 
dispersion relation, but no other model dependent assumptions are made.
In particular there are no spurious parameters
included in the IAM derivation, but just the ChPT LECs, $m_\pi$ and $f_\pi$.

Remarkably, the simple Eqs.\eqref{mIAM} (either the IAM or the mIAM) 
ensure elastic unitarity, match ChPT amplitudes
at low energies, and,  using LECs 
compatible with existing determinations,
describe fairly well data up to somewhat 
less than 1 GeV, generating
the $\rho$, $K^ *$, $\sigma$ and $\kappa$ resonances
as poles on the
second Riemann sheet.

Of course,
other unitarization techniques are possible, but in order to improve
the NNLO IAM these would imply
the use of coupled channels,
 which are not needed for the $\sigma$  and the $\rho$, 
higher orders of ChPT, which, if available, 
could be incorporated in a higher order IAM version, 
or a different approximation of the left cut.
For the latter, we can distinguish two possible regions for
improvements:
either at low energies, where we could systematically
 include more orders of ChPT, 
again leading to a higher order IAM version, or
at high energies (which are nevertheless suppressed by 
the subtractions) where the left cut should be modeled
introducing more - non ChPT -- parameters whose dependence on QCD is not
known, or further assumptions beyond those used in the IAM.
There is also, of course, an ambiguity on where to 
choose the subtraction point, but, as shown in \cite{GomezNicola:2007qj}, 
different choices contribute to higher order corrections in the subtraction
constants and the pole contribution terms, which numerically
differ very little from one another.

Finally, there are other unitarization techniques which are very successful
and simpler than the IAM. However these can be recast, in the elastic regime,
as the IAM plus further approximations, like dropping 
crossed and tadpole terms \cite{Oller:1997ng} -- and therefore keeping a 
spurious parameter like a cutoff or an unknown 
subtraction constant to regulate the theory --
or keeping just the leading order \cite{Oller:1997ti}, 
in which case the $\rho$ 
cannot be reproduced with natural size parameters (see discussion below). 
Some of these simpler methods can be easily understood in terms
of resummations of particular sets of Feynmann diagrams, but
without the said additional simplifications 
there is no proof that the IAM, and much less so the mIAM, can be obtained from a simple diagrammatic
resummation.
All these simpler methods are known to provide  physical
results for the scalars rather similar to the IAM. 
However, in order to extrapolate
to non-physical quark masses one would need additional model dependent
assumptions on the behavior of the spurious parameters.
For the reasons explained in the last two paragraphs 
the IAM is the most adequate and complete technique to study the quark
mass dependence of the $\rho$ and $\sigma$ elastic resonances.

For our purposes in sec.\ref {sec:op6mIAM}
it is important to remark that it has been shown \cite{Oller:1997ti} 
that the scalars can actually be generated mimicking
the LEC, tadpole and crossed channel diagrams by a cutoff of natural size,
and thus it is said that scalars are ``dynamically generated'' from,
essentially, meson-meson dynamics (meson loops). In contrast,
to generate the vectors, {\it a precise knowledge of the LECs is needed}, 
namely, of the underlying, non meson-meson QCD dynamics. As we will see
this makes the $\rho(770)$ much more sensitive to the still poorly known
two-loop LECs, whereas the sigma is rather stable under such higher order corrections.

Thus, by changing $m_\pi$ in the IAM amplitudes we can study how the 
generated $\rho$ and $\sigma$ poles evolve, so
we can predict the dependence of their masses, widths and
couplings on $m_\pi$. The values of $m_\pi$ to be considered should
lie in the applicability region of ChPT and should allow for 
some $\pi\pi$ elastic regime. Both criteria would fail 
above, roughly, $m_\pi\simeq 500$ MeV, which was taken as the upper
bound for one-loop calculations in \cite{chiralexIAM}.
We will see here that the approach is not reliable much before,
namely, around $m_\pi\simeq 300-350\,$ MeV at least.

In \cite{chiralexIAM}
the mIAM was used for the $\rho$ and $\sigma$ chiral extrapolation,
because, for the scalar and  at high $m_\pi$,
one resonance pole gets near the IAM spurious pole,
a problem that is nicely solved with the mIAM. Nevertheless,
in the physical region and near the other generated poles, the
differences between IAM and mIAM approaches are almost negligible, even for
high pion masses. 
What we have briefly reviewed is just the NLO, or one-loop, case
derived in \cite{GomezNicola:2007qj}. For this work we will
need the two-loop version of the mIAM that we will explicitly calculate 
in sect.\ref{sec:op6analysis} below.
But first we will show that the predictions obtained in \cite{chiralexIAM}
for the chiral extrapolation of the one-loop case are in quite
good agreement with recent lattice results on the pion
mass dependence of the $\rho$ mass,
$f_\pi$ and isospin 2 scattering length.

\section{One-loop results}
\label{sec:results}

Within the SU(2) ChPT formalism, for $\pi\pi$ scattering 
only four LECs appear at $O(p^4)$, and are denoted by $l^r_1,\cdots l_4^r$.
Since for now  we just want to compare the predictions in \cite{chiralexIAM}
with lattice results, we will use  the same values as in 
\cite{chiralexIAM}. Namely, we use 
$10^3 l_3^r=0.8\pm 3.8$, $10^3 l_4^r=6.2 \pm 5.7$ from 
\cite{Gasser:1983yg}; and $l_1^r$ and
$l_2^r$ are obtained from a mIAM fit to phase shift
data up to the resonance region, $10^3 l_1^r=-3.7\pm 0.2$,
$10^3l_2^r=5.0\pm 0.4$. All the LECs are evaluated at $\mu=770$ MeV.

Let us first compare with lattice results and then we will calculate 
explicitly the coupling constant of resonances to two pions.

\subsection{Comparison with lattice}
\label{sec:comparisonwithlattice}

Around the time of the publication in \cite{chiralexIAM},
several lattice
calculations were published providing results 
for the pion decay constant $f_\pi$ 
\cite{lattice2,fpilattJLQCD,Beane:2007xs} and also for the $\pi\pi$ 
scalar isospin 2 scattering length $a_{20}$
\cite{Beane:2007xs}. What we will show next is that, despite these lattice results
were not used as input in the calculations
of \cite{chiralexIAM}, they are fairly 
well described within our one-loop formalism.

\begin{figure}[t]
  \centering
%  \hbox{
    \includegraphics[angle=-90,scale=.5]{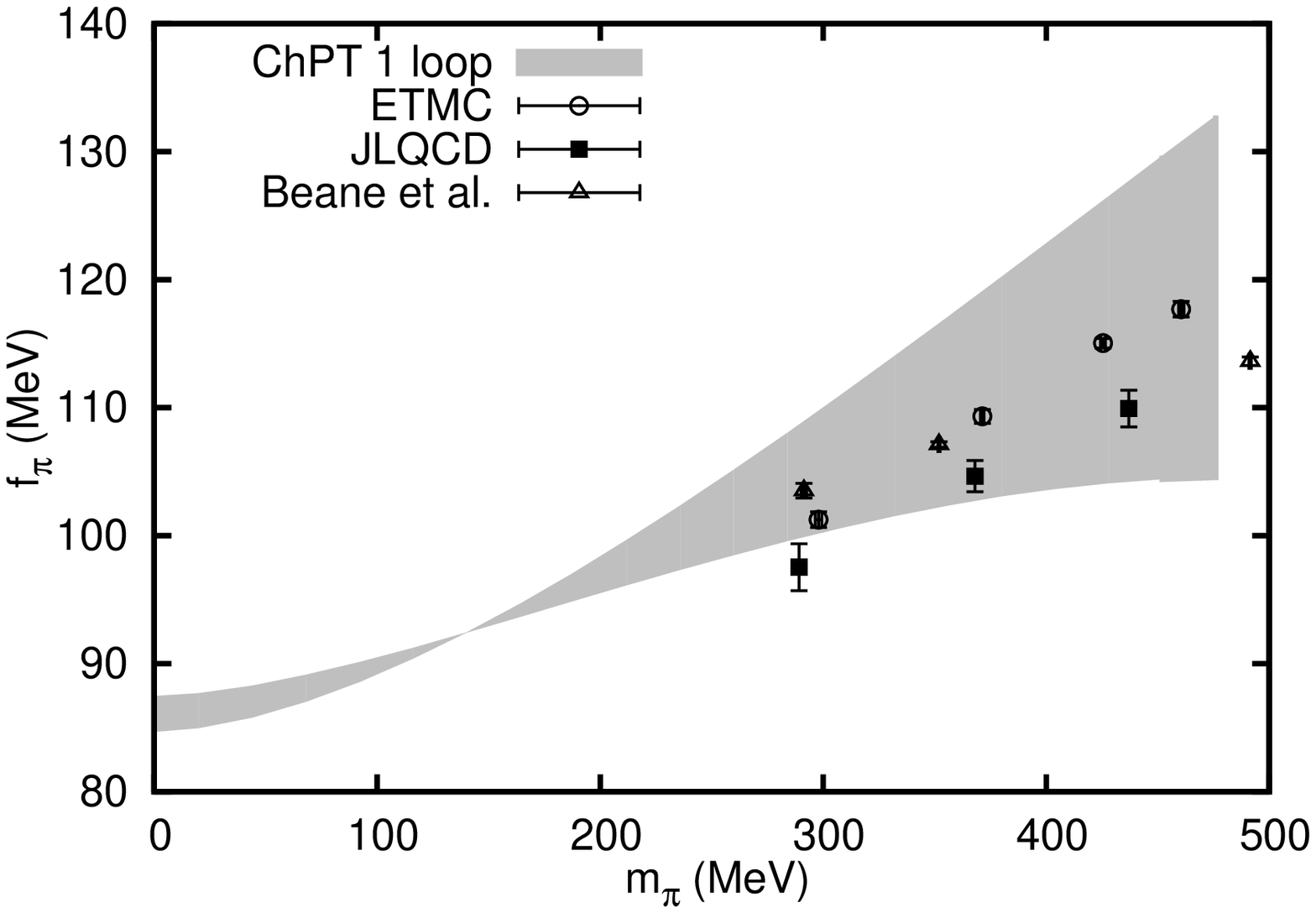}\\
    \includegraphics[angle=-90,scale=.5]{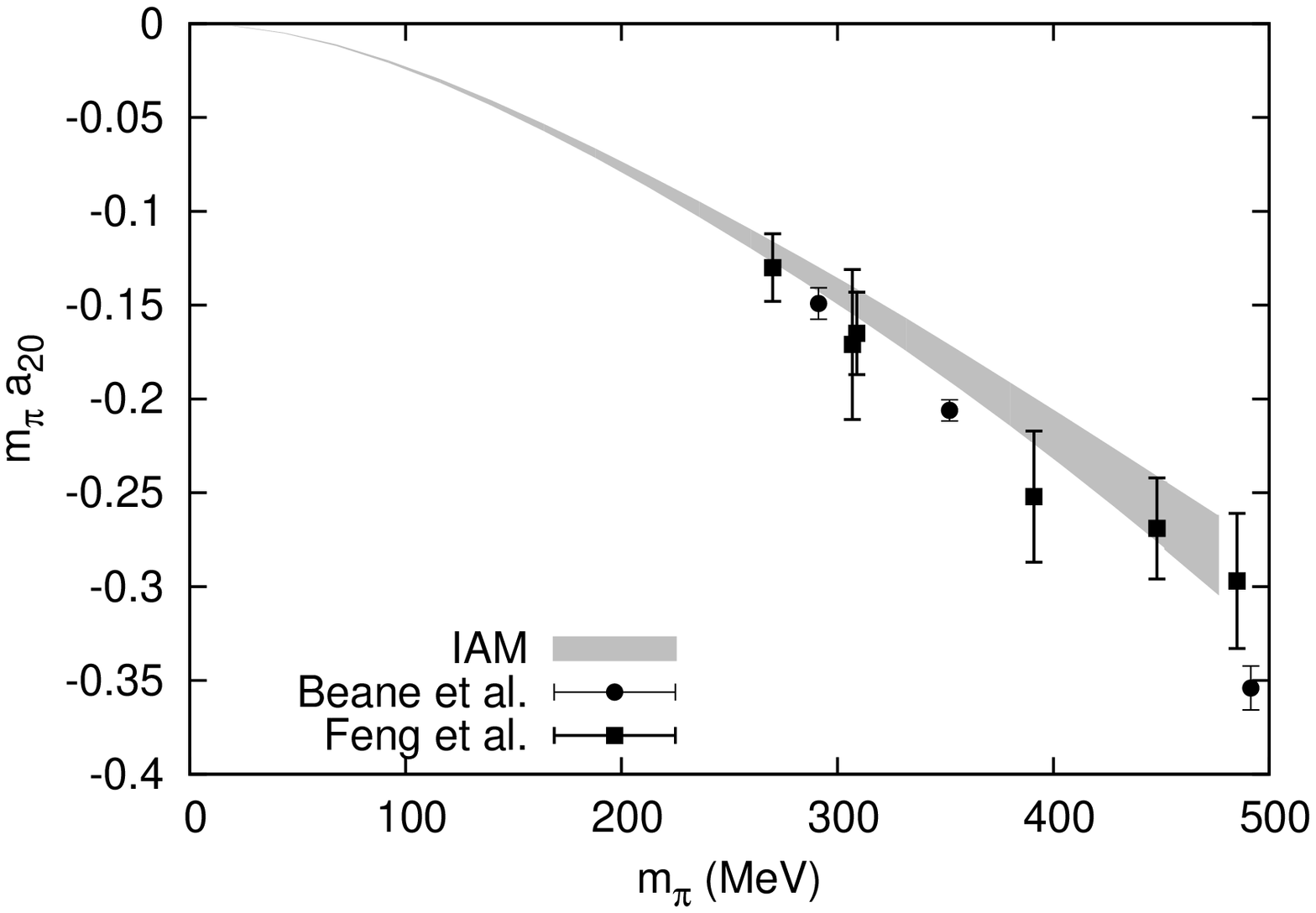}
%  }
  \caption{Top: One loop
  ChPT and lattice results for the $f_\pi$ dependence on $m_\pi$,
but with the $l_4^r$ parameter used in our IAM fits.
  Lattice points are from 
  \cite{lattice2,fpilattJLQCD,Beane:2007xs}.
Bottom: IAM and lattice results for the
  $I=2,J=0$ scattering length $a_{20}$ dependence on $m_\pi$.
  Lattice points are from \cite{Beane:2007xs,Feng:2009ij}. 
}
  \label{a20fpi}
\end{figure}

In particular, when changing $m_\pi$ in our amplitudes we have to change
accordingly the value of $f_\pi$, as it also depends on $m_\pi$.
Note that, of course, the $f_\pi$ calculation is not unitarized, 
but just standard ChPT. However, $f_\pi$ is an important factor 
in all our unitarized calculations.   
Actually, since we are using ChPT up to $O(p^4)$ 
in the IAM dispersion relations,
 we evaluate the 
$f_\pi$ dependence on $m_\pi$ to that order. At $O(p^4)$,
the only LEC that appears in the $m_\pi$ dependence of $f_\pi$
is $l_4^r$, whose value is fixed here to that given in
\cite{Gasser:1983yg}, as commented above. 
In the top panel of Fig.~\ref{a20fpi}
we compare the resulting one loop dependence of $f_\pi$ on $m_\pi$ with
some lattice results. The grey area covers the uncertainty in $l_4^r$
only,
and one should also recall that lattice uncertainty bars are statistical.
We consider the spread of the different lattice groups as an estimate of their
systematic errors. With these remarks in mind, 
we can see that the $f_\pi$ dependence implemented
in our approach is compatible with that calculated from the lattice.
Our grey band does not extend beyond $m_\pi=0.5\,$ GeV because, 
at the very least,
 that is for sure an applicability bound for our calculations. 
In the next sections,
we will see that at those pion masses the two-loop uncertainties are
actually too big to make any significant quantitative claim with our
method, which we will find reliable only up to 
roughly $m_\pi\simeq 300-350\,$ MeV, at most.

The $I=2, J=0$ $\pi\pi$ partial wave can be calculated with the IAM,
and indeed it was included in the fit to phase shift data \cite{chiralexIAM},
thus constraining its energy dependence.
However, the $m_\pi$ dependence of $a_{20}$ was not constrained with any input.  
Of course, it can be easily predicted and is interesting to compare it
with the available lattice data as a 
consistency check of the method. Thus, in the bottom panel of
Fig.~\ref{a20fpi} we show the lattice results for $a_{20}$
compared to our IAM calculation.
Once again, one should take into account that the mIAM  error 
band only covers the uncertainties in the one-loop LECs. 
In view of the figure, and taking into account that our curve is not a fit to these data,
we consider that our predictions are in fairly 
good agreement with these lattice results.

Of course, we could refit our approach including the data in Fig.\ref{a20fpi},
and we would be getting a better agreement, but at this point 
we only want to check the consistency of our results
and that we do not need a fine tuning to describe lattice results,
which will be included as input of two-loop fits in the next sections below.

We have thus checked our method
for consistency against available lattice results on two quantities
other than resonance masses, which we address now.
First, let us recall that
the mass $M$ and width $\Gamma$ of a narrow resonance are related to its
pole position as $\sqrt{s_{pole}}=M-i\Gamma/2$, and
this notation is usually also kept for wide resonances,
as done in \cite{chiralexIAM}.
However, lattice calculations do not provide results in the complex plane.
For that reason, in this work
we will also consider for the $\rho(770)$ the most usual and physically
intuitive definition: namely, that the mass of the resonance corresponds to the energy
where the scattering phase shift reaches $\pi/2$.
This is the value where the modulus of the scattering amplitude shows a peak,
and we will thus call it ``peak mass''. 
Of course, this definition is only
valid for narrow resonances, and is a very good approximation for the $\rho$,
which is the one for which more reliable lattice results exist. For the $\sigma$ 
we will stick to the pole mass definition.

\begin{figure}[t]
  \centering
  \includegraphics[scale=.5,angle=-90]{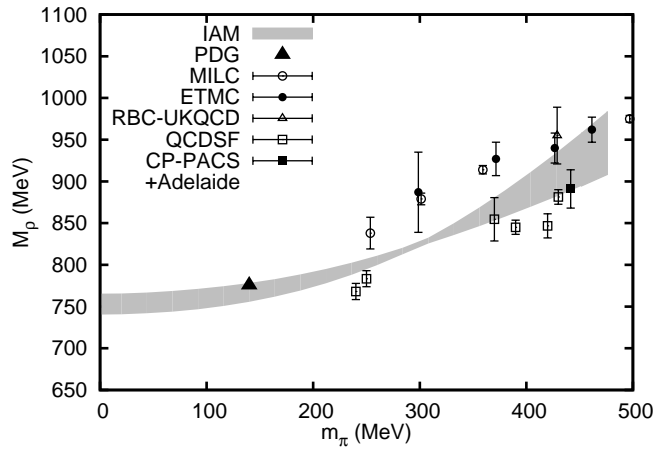}
  \caption{Comparison of our IAM results for the $M_\rho$ 
    (calculated  here as the point where the phase 
    crosses $\pi/2$) dependence on
    $m_\pi$ with some recent lattice results 
    \cite{lattice1,lattice2,lattice3,lattice4,lattice5}. The
    grey band covers only the error coming from the LECs uncertainties.}
  \label{Mrho}
\end{figure}

Thus, in  Fig.~\ref{Mrho} we show the results of 
 our IAM calculation of the $\rho$ mass 
evolution when $m_\pi$ varies,  which is displayed as a 
grey band that covers the uncertainties in the one-loop fitted LECs only.
This behavior  agrees nicely with the estimations 
for the two first coefficients of the $M_\rho$  chiral expansion \cite{bruns}.
This figure is relatively similar to Fig.~3 in \cite{chiralexIAM},
except that we plot versus $m_\pi$ and not $m_\pi^2$ and that
we have defined the mass as the energy where the phase shift crosses $\pi/2$.
This latter choice ensures that the physical $\rho$ mass from the PDG 
\cite{PDG} lies
within our uncertainty band. In contrast, the ``pole'' mass would lie somewhat lower. This difference between ``pole'' and ``peak'' masses decreases
as $m_\pi$ increases, and they are almost indistinguishable around 350 MeV.
Of course, we are now comparing with a compilation of
lattice results from different collaborations.
Let us remark  that due to the finite lattice volume, for some of the lattice results, the minimum energy with which pions are produced is larger than the resulting $M_\rho$ and therefore the resonance has zero width.
With these caveats in mind, 
and in view of the large systematic deviations between different 
lattice collaborations, one can conclude that our one-loop 
prediction shows a rather 
good agreement with the bulk of lattice results.

\subsection{Resonance couplings}
\label{sec:resonancecouplings}

In Ref. \cite{chiralexIAM} it was shown that both
the $\rho$ and $\sigma$ widths -- calculated with the mIAM from the
imaginary parts of the pole positions--
decrease as $m_\pi$ is increased. 
It was also shown that
the decrease in $\Gamma_\rho$ is largely
kinematical, following remarkably well the expected
reduction from phase space as $M_\rho$ approaches threshold,
whereas the $\sigma$ width decreases in a quite different way
of that provided only by phase space reduction.
Following this argument it was concluded in \cite{chiralexIAM}
that the effective coupling of the $\rho$ to $\pi\pi$ must be
almost $m_\pi$ independent and that the $\sigma$
coupling should show a strong $m_\pi$ dependence.
However, we did not perform an
explicit calculation of these 
couplings, and we will provide it here. 
For elastic amplitudes with a given isospin $I$ and angular
momentum $J$, we can define their coupling $g$ 
to two pions as
\begin{equation}
  \label{geff}
  g^2 \sim \lim_{s\to s_{p}}\frac{(s-s_{p})t_{IJ}(s)}{p^{2J}}, 
\end{equation}
where $s_p$ is the position of the resonance pole on the second Riemann sheet,
and $p$ is the center of mass momentum.
The above definition, if used as such, 
is very unstable numerically. For that reason
 we have calculated the residue applying Cauchy's Theorem to a small circle
surrounding the pole.

\begin{figure}[t]
  \centering
  \includegraphics[angle=-90,scale=.5]{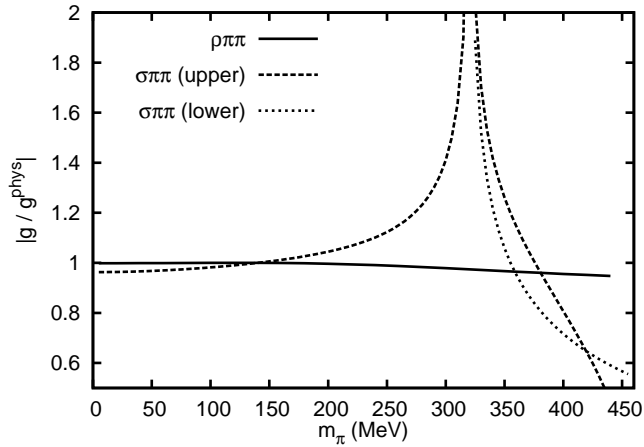}
  \caption{$\rho$ (continuous line) and $\sigma$ (dashed line)
  couplings to $\pi\pi$ evolution with $m_\pi$. Since the two conjugate
$\sigma$ poles become two real 
poles after reaching the real axis, there is a coupling for the 
pole that moves towards
 threshold (``up'') and another one for the pole moving away (``down'').}
  \label{fig:couplings}
\end{figure}

In  Fig.~\ref{fig:couplings} we show the results for the couplings.
The continuous line shows the $\rho\pi\pi$
effective coupling evolution as $m_\pi$ changes, defined as in
Eq.\eqref{geff} from the IAM partial wave 
and normalized to its physical value. As
expected, it is almost $m_\pi$ independent,
since it deviates by less than 5\% from its original value
when changing the pion mass from 139.57 MeV up to 450 MeV. 

The results of our explicit calculation thus confirm quantitatively
the $\rho$ coupling independence 
suggested qualitatively in  \cite{chiralexIAM}.
This result is very relevant because it justifies
the constancy assumption made in lattice studies
of the $\rho(770)$ width \cite{Aoki:2007rd}.

In contrast, the $\sigma\pi\pi$ coupling, plotted as
the dotted line of Fig.~\ref{fig:couplings} shows a
strong $m_\pi$ dependence. The dramatic peak in this curve, 
around 320 MeV corresponds to the 
pion mass where, as shown in \cite{chiralexIAM},
 the two conjugated poles of the sigma meson  on the second Riemann sheet
join on the real axis below threshold. As the pion mass increases 
beyond that value, the
two poles remain real and we therefore have to plot two different branches.
One of these two poles (labeled ``lower'')
stays away from threshold, at least  within the mass range of this study.
In contrast, the other one (labeled ``upper'') 
moves towards threshold 
and, eventually
jumps onto the first Riemann sheet. 
For the one loop calculation this occurs slightly above $m_\pi=450\,$MeV,
but we will see that for the two-loop calculations this may occur 
for pion masses even only slightly above 300 MeV.
Let us also remark that it is a nice consistency check 
to see that, as the pole reaches threshold,  the coupling tends to zero, as
it is expected from general arguments \cite{Weinberg:1962hj,Gamermann:2009uq}.

As it could be expected, our results both for the $\rho(770)$ and the $\sigma$
 are also in very good agreement 
with a very recent one-loop calculation using unitarized 
SU(3) ChPT \cite{Nebreda:2010wv}, 
since the strange quark mass plays a
 very little role in the $\rho(770)$ and $\sigma$ 
values of their masses and widths. 
Of course, we expect our SU(2) results to be more reliable than those of
SU(3) since, as it is well known, the SU(2) ChPT convergence is much better.
The small differences with 
respect to \cite{Nebreda:2010wv}
can only be attributed to small differences in the choice of LECs. Actually
the SU(2) LECs that  we use here were fitted in
\cite{chiralexIAM} 
to pion-pion scattering data only. In contrast, in  \cite{Nebreda:2010wv}
the SU(3) LECS used there were fitted to the whole elastic meson-meson
scattering data below 1 GeV plus some lattice results. 
The effect of the the kaon loops, which have been 
integrated out in the large kaon mass limit, is negligible compared to our uncertainties.

\section{Two-loop formalism and results}
\label{sec:op6analysis}

In this section we extend our analysis to the next to next to leading order (NNLO), 
namely, two-loop unitarized ChPT. As we will see, the present knowledge about
the two-loop LECs is rather poor. However, it will be enough to obtain sufficiently robust predictions for some observables, particularly those related to the $\sigma$ meson, and to 
estimate the size of the uncertainties due to higher order terms
in the different parts of the calculation. 
First of all, of course, we have to rederive 
the mIAM formalism for the two-loop case.

\subsection{The two-loop mIAM formalism}
\label{sec:op6mIAM}

As we have already commented, instead of the IAM, we need the mIAM 
in order to describe properly
the subthreshold region in the scalar waves. 
As explained in subsection \ref{sec:IAM}, in the dispersion integral 
of Eq.\eqref{disp1/t}
 we now have to evaluate the left cut and the
pole contribution using
the $O(p^6)$ expansion of the ChPT amplitudes: $t=t_2+t_4+t_6+...$. 
Of course, now the position of the Adler zero 
has to be evaluated also up to that order, i.e., $s_A\simeq s_2+s_4+s_6$.
After a tedious calculation along the lines of 
\cite{GomezNicola:2007qj}, we arrive at the $O(p^6)$ version of the mIAM:
\begin{equation}
  \begin{aligned}
    t^{mIAM}&=\frac{t_2^2}{t_2-t_4+t_4^2/t_2-t_6+A^{mIAM}},\\
    A^{mIAM}&=t_4(s_2)-\frac{2t_4(s_2)t'_4(s_2)}{t'_2(s_2)}
    -\frac{t_4^2(s_2)}{t'_2(s_2)(s-s_2)}\\
    &+t_6(s_2)+\frac{(s-s_2)(s_A-s_2)}{s-s_A}
    \Bigg(
      t'_2(s_2)-t'_4(s_2)\Bigg.\\
      &\Bigg.-t'_6(s_2)+\frac{t'_4(s_2)^2+t''_4(s_2)t_4(s_2)}{t'_2(s_2)}
    \Bigg),\label{mIAMp6}
  \end{aligned}
\end{equation}
As a technical remark, let us note that
we have to introduce an additional subtraction to ensure convergence.
Note also that we will now need the second derivative of $t_4$ with respect to $s$
at $s_2$.
Again, the standard two-loop IAM \cite{Dobado:1996ps,Nieves:2001de} is recovered for $A^{mIAM}=0$, 
which occurs for all waves with $J>0$.
We can now use the $O(p^6)$ mIAM to calculate the chiral extrapolation
of the $\rho$ and $\sigma$ resonance poles, and check the
consistency with the $O(p^4)$ results. 

\subsection{Two-loop ChPT and the low energy constants}
\label{sec:op6LECS}

The two-loop $\pi\pi$ SU(2) ChPT scattering amplitude 
was calculated in \cite{Bijnens:1997vq} and contains 
six additional LECs. These are denoted 
$r_1,\cdots r_6$ and their values are poorly known. 
Moreover, this $O(p^6)$ calculation  is expanded
in terms of $(m_\pi/f_\pi)^2$, where $f_\pi$ is the pion decay constant
evaluated at the physical value of the pion mass, not in the chiral limit. 
That is fair enough to describe the physical scattering amplitude,
but in order to extract the pion mass dependence of the scattering amplitude
one also has to include the  $O(p^6)$ ChPT $f_\pi$ pion mass dependence
\cite{Bijnens:2006zp}:
\begin{eqnarray}
\frac{f_\pi}{f_0}&=&
1+\frac{m_\pi^2}{f_0^2}\left(l_4^r-L\right)+\frac{m_\pi^4}{f_0^4}
\left[-\frac1{N}
  \left(
    \frac{l_1^r}{2}+l_2^r
  \right) \right. \nonumber\\
&&+ 
  L\left(7l_1^r+2l_2^r-l_4^r+\frac{29}{12 N}\right)   
-\frac{3}{4}L^2  \nonumber\\
&& 
\left.
 -2l_3^rl_4^r + \frac1{N^2} 
  \left(
    \tilde r_f - \frac{13}{192}
  \right)
\right],
\end{eqnarray}
where $L=\frac1{N}\log\left(m_\pi^2/\mu^2\right)$, $N=16\pi^2$ and 
$\tilde r_f$ is the relevant combination of $O(p^6)$ LECs that
appears in $f_\pi$, poorly known once again. For that reason, we will
use lattice data on $f_\pi$ to stabilize it in our fits.

Let us now remark that the two-loop
leading log contributions, which are numerically dominant at low energies,
do not depend on the $r_i$ constants, but just on the one-loop $l_i^r$.
For this reason
it is well known that the  values of the $l_1^r\cdots l_4^r$ $O(p^4)$ LECS 
can vary sizably between the $O(p^4)$ and $O(p^6)$ 
analysis, as it is shown in Table \ref{LECsOp4}, where we quote
the LECs obtained in several works 
\cite{Girlanda:1997ed,Amoros:2001cp,Buettiker:2003pp,Colangelo:2001df}. Let us remark that error bars in the
$l_1^r\cdots l_4^r$
usually correspond
to the statistical uncertainty (``noise'') 
in the input, but there are other large systematic sources of uncertainty
that are most likely dominant. 
Hence, it should not be surprising that
some of the $O(p^6)$ LECs deviate somewhat from the estimates
when taking into account leading log terms from $O(p^8)$ and higher orders,
as we will do with the IAM.

\begin{table}
  \caption{Sample of LECs:
First row: Roy-Steiner eqs. SU(3) analysis of $\pi K$ scattering.
Second  and third rows: $K_{l4}$ analysis to $O(p^4)$ and 
$O(p^6)$, respectively.  Naively, we have 
combined quadratically the SU(3) LECs errors there. 
Fourth row: Roy Eqs. analysis with 
uncertainties from imaginary parts and unknown 
$O(p^6)$ LECs combined quadratically.
Fifth row: Roy Eq. analysis of low-energy 
$\pi\pi$ scattering up to two loops, whose errors
 ``only account for the noise seen in their calculations''.
Last row, values used in \cite{chiralexIAM} with the IAM.
    All LECs are evaluated at the scale $\mu=770$ MeV
}
  \label{LECsOp4}
  \begin{tabular}{|c|c|c|c|c|}
    \hline
    \vrule height 10pt depth 5pt width0pt
    Analysis & $10^3l_1^r$ & $10^3l_2^r$ & $10^3l_3^r$ & $10^3l_4^r$ \\
    \hline
    $O(p^4)$ \cite{Buettiker:2003pp} &$-4.9\pm0.6$&$5.2\pm0.1$& -- &$17\pm10$ \\
    $O(p^4)$ \cite{Amoros:2001cp} &-4.5&5.9&2.1&5.7 \\
    $O(p^6)$ \cite{Amoros:2001cp}&$-3.3\pm2.5$&$2.8\pm1.1$&$1.2\pm1.7$&$3.5\pm0.6$ \\
    $O(p^6)$ \cite{Girlanda:1997ed} &$-4.0\pm2.1$&$1.6\pm1.0$&--&  --\\
    $O(p^6)$ \cite{Colangelo:2001df}&$-4.0\pm0.6$&$1.9\pm0.2$&$0.8\pm3.8$&$6.2\pm1.3$ \\
\hline
IAM \cite{chiralexIAM}&$-3.7\pm0.2$&$5.0\pm0.4$&$0.8\pm3.8$&$6.2\pm5.7$\\
    \hline
  \end{tabular}
\end{table}

Finally, it has been shown \cite{Ecker:1988te}
that the LECs can be understood as the
effective couplings that
result from integrating out heavy fields.
Indeed, most one-loop LECs values are saturated by vector
resonance exchange, like the $\rho(770)$, but note that the $\sigma$
plays a little role, if any, in the actual values of the LECs.
These resonance saturation estimates have also been extended
to the $O(p^6)$ LECs \cite{Bijnens:1997vq}, but 
these are very uncertain and are customarily assigned a 100\% uncertainty.
We have collected them in Table \ref{LECs}, together with some other estimates 
coming from dispersive analysis of pion-pion scattering \cite{Colangelo:2001df}.

At this point we want to make clear that the LECs do {\it not}
depend on the quark mass. 
This is a rather obvious statement for people familiar with ChPT but,
when presenting in conferences and workshops \cite{Pelaez:2010er,conferences} previous results 
from  Ref.~\cite{chiralexIAM} or partial preliminary results from this work,
we have found that people get confused since we have just stated that
the $l_i^r$ are saturated by resonance exchange -- like $\rho$ exchange --
but we have shown that the $\rho$ mass depends on the quark mass.  
The reason for this confusion is that
resonance saturation is usually 
interpreted as a $\sim 1/M_R^2$ contribution to the $l_i^r$, with $M_R$ the physical mass of a resonance. However  
it is actually a $\sim1/M_{R0}^2$ contribution, 
with $M_{R0}$ the resonance mass in the chiral limit,
namely $M_R^2=M_{R0}^2+O(m_\pi^2)$.
Numerically, using $M_R$ instead of $M_{R0}$
 makes a small difference --
neglected when obtaining LECs estimations --
but is incorrect in terms of ChPT.  
A term like  $\sim 1/M_R^2$ coming from integrating out a 
heavy resonance  should be re-expanded as 
$\sim1/M_{R0}^2(1+ O(m_\pi^2/M_{R0}^2))$.  The first term
contributes to $l_i^r$, but the next $O(m_\pi^2)$ term counts as  a higher 
order in ChPT and therefore does not contribute to the same order 
as $l_i$ does.
The same occurs to all orders, so, as stated above, {\it the LECs do not have to be readjusted
when the mass of higher resonances change with $m_\pi$. They do not depend on the quark mass.}

\subsection{Resonance sensitivity to LECs}
\label{Ressens}

After reviewing briefly the standard ChPT two-loop calculation, 
we can now use it to generate the $\rho(770)$ and $f_0(600)$ resonances, for which
we use the mIAM. Thus, we show once again in Table \ref{LECsOp4}
the LECs we obtained in the 
one-loop IAM fit in \cite{chiralexIAM}, which are fairly compatible 
with the non-unitarized
determinations, lying roughly 
in between the one and the two-loop bulk determinations.
Naively, this could correspond to the fact
that the one-loop IAM,
reproduces not only the one-loop ChPT expansion
but also the numerically relevant s-channel two-loop diagrams.

At this point it is important to recall the different
role that LECs play in the generation of the $\rho(770)$ and the $\sigma$
resonances. In particular, 
it has been shown that the $\sigma$ can be easily generated
within the chiral unitary approach \cite{Oller:1997ti} from the leading
order ChPT -- which only depends on $f_\pi$ and $m_\pi$ -- 
and a natural cutoff, whereas that is not feasible for the
$\rho(770)$, which needs the input from the one-loop LECs 
\cite{Oller:1997ng}. 
This is also understood from the $1/N_c$ behavior of these resonances
\cite{Pelaez:2003dy,Pelaez:2006nj}:
namely, the $\rho(770)$ behaves nicely as a $\bar qq$ state when
generated from the IAM -- a $N_c$ dependence due to the
the $1/N_c$ behavior
of the leading LECs. In contrast,
the $\sigma$ does not behave {\it predominantly}
 as a $\bar qq$ and this dependence is mostly
 due to logarithmic terms,
which are independent of the LECs and
are generated from meson loops. This explains why the $\sigma$,
despite being lighter than the $\rho$,
is not contributing so sizably to the value of the low energy constants.
In summary, 
the terms containing LECs play a crucial role in
the generation and location of the $\rho$ pole in the IAM, but not
so much for the $\sigma$.
This is due to the fact that to generate the $\rho$ 
we need the input from the 
underlying QCD dynamics of a $\bar qq$ state, and we expect
this to occur to all orders in ChPT, whereas
the sigma
is dominated by the scale $f_\pi$ \cite{Oller:1997ti} and should depend much more mildly on the
underlying QCD dynamics encoded in the LECs. 

This is actually what we find when we look for the $m_\pi$
dependence of the $\rho$ pole with the $O(p^6)$ IAM;
its behavior is quite unstable under the large uncertainties
of the $O(p^6)$ LECs.  If we leave completely free the $O(p^6)$
parameters within their huge  estimated uncertainties, we cannot make
any two-loop prediction for the $\rho(770)$. For this reason, at two loops
 we will fit not only experimental data but also the LECs values.
In addition, by fitting the 
experimental data only, as was done in \cite{chiralexIAM},
 one constrains mostly the combinations of $O(p^6)$
chiral parameters that govern the energy dependence. 
However, one cannot expect to constrain the $O(p^6)$ LECs 
combinations that govern the pion mass
dependence that is of interest here.
Fortunately, as we have seen in previous sections, there is a large amount of
lattice data on the pion mass dependence of the
$\rho(770)$ mass, the $I=2$ scattering length and
$f_\pi$, which can be used to constrain further the $O(p^6)$ LECs
and will be included in our fits.
Once this is done, we will obtain predictions for the  $m_\pi$ dependence
of the $\rho$ coupling and width, as well as on
all the $f_0(600)$ parameters,
which is where the most interesting discussion is still going on.

\subsection{Unitarized two-loop constrained fits}
\label{sec:op6fits}

\begin{figure*}
  \centering
  \vbox{
    \hbox{
      \includegraphics[scale=.48,angle=-90]{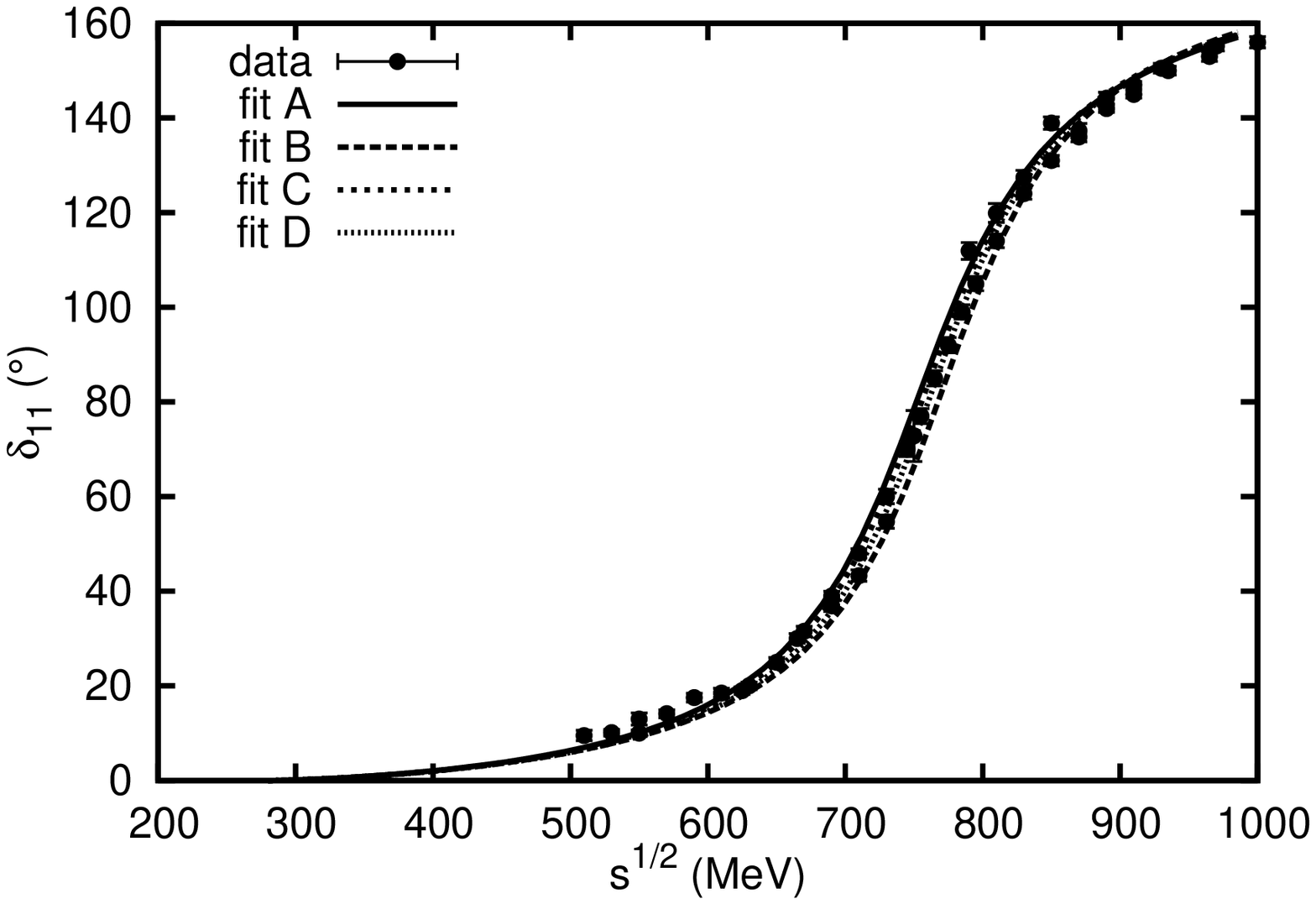}
      \includegraphics[scale=.48,angle=-90]{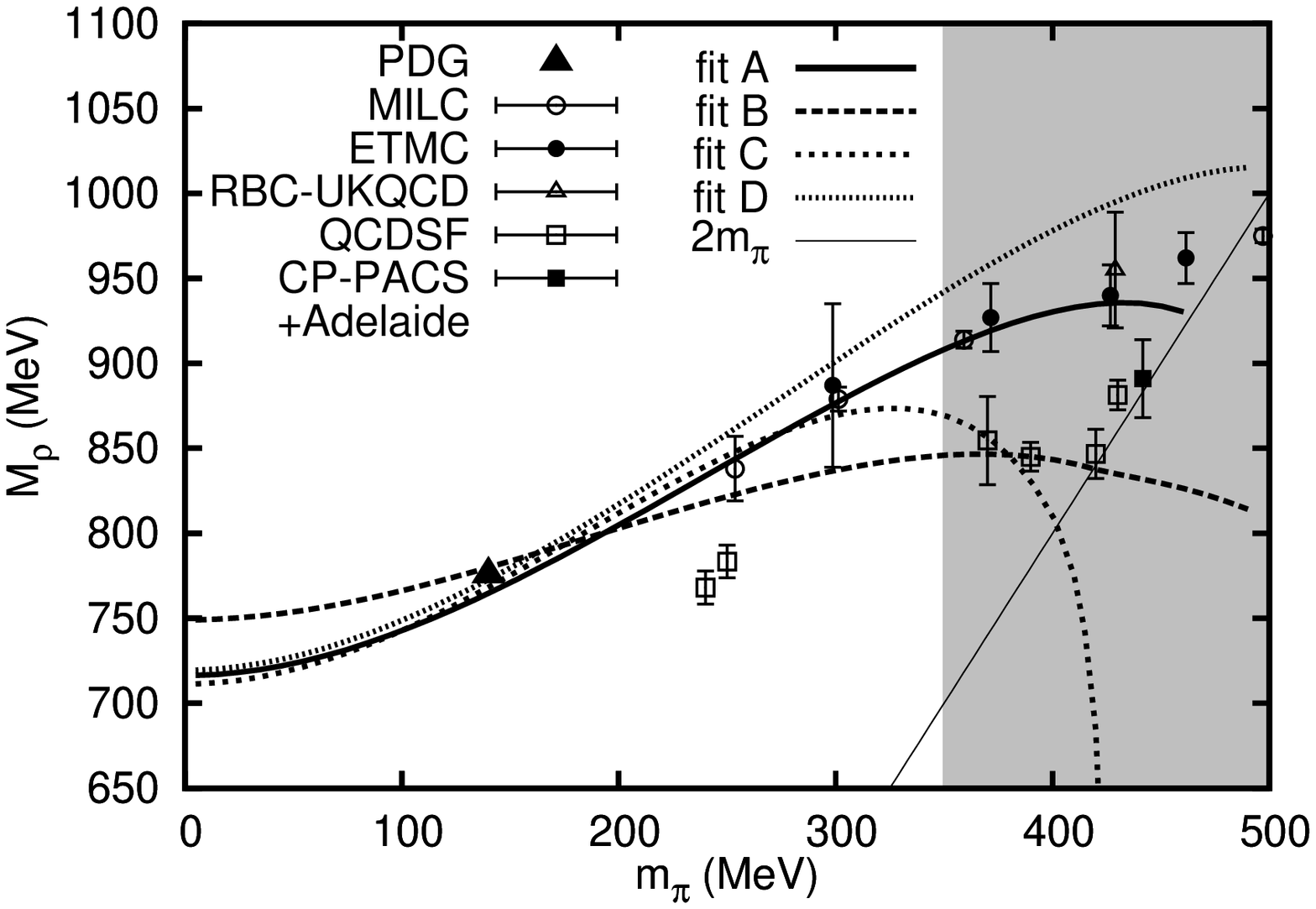}
    }
    \hbox{
      \includegraphics[scale=.48,angle=-90]{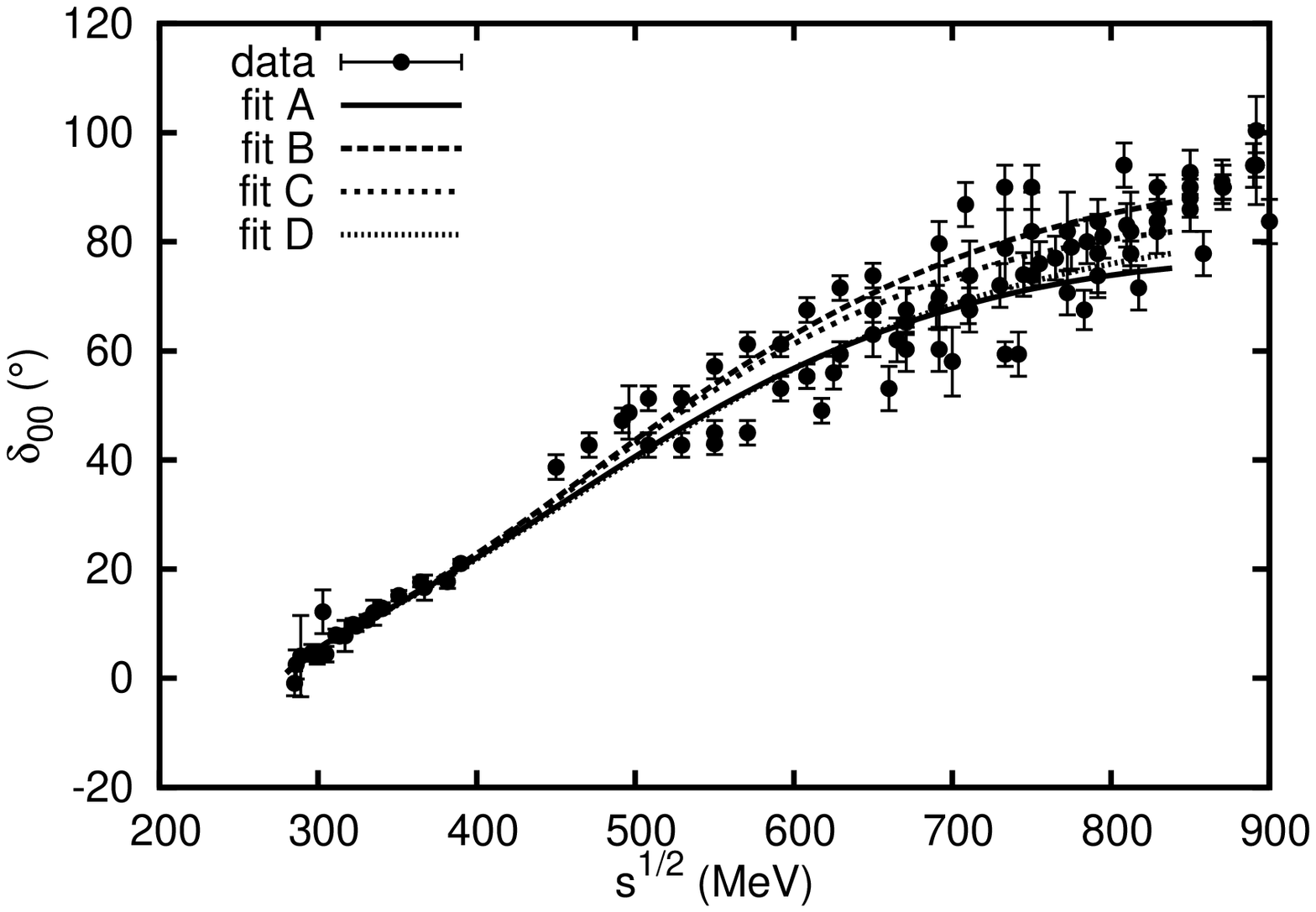}
      \includegraphics[scale=.48,angle=-90]{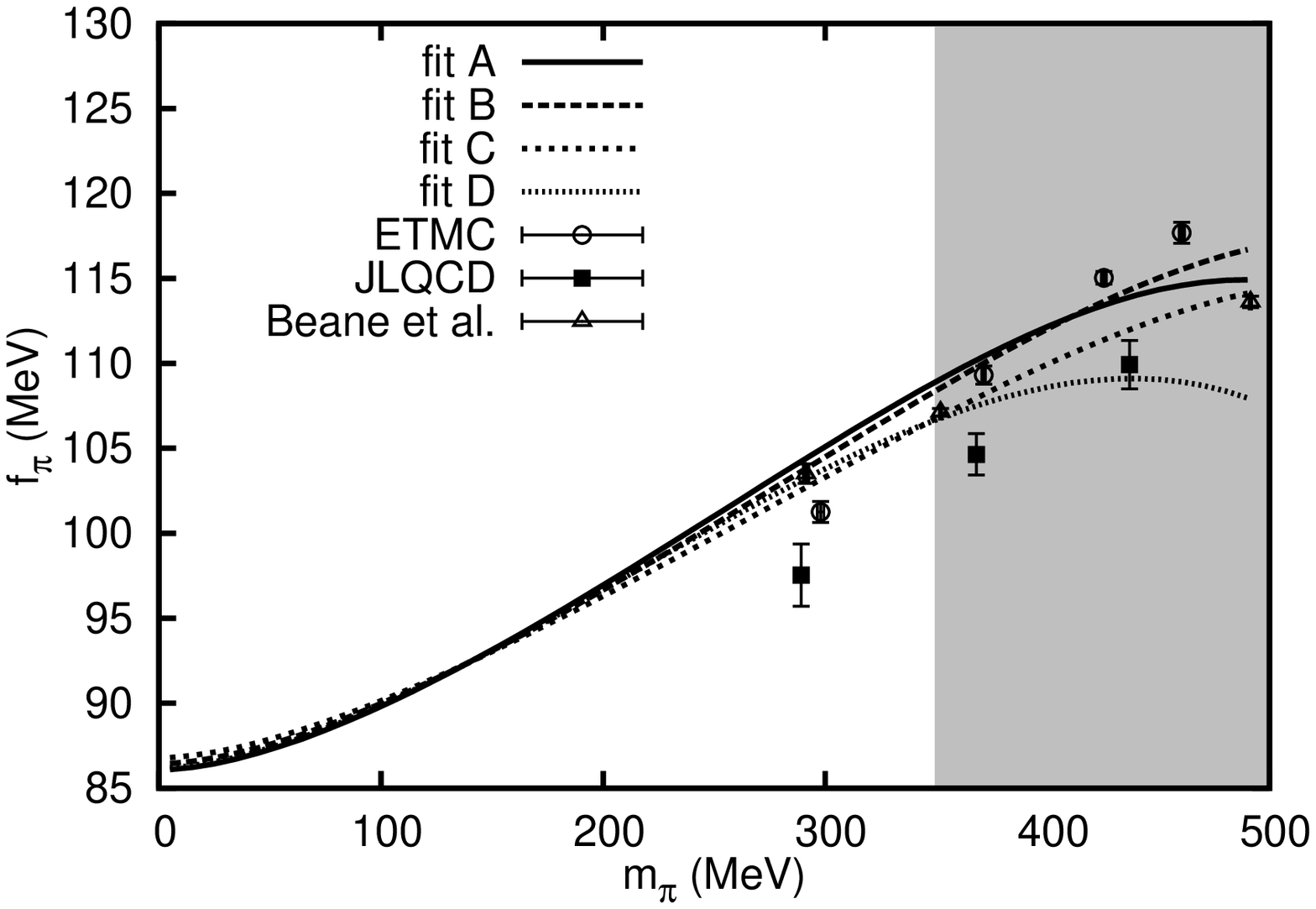}
    }
    \hbox{
      \includegraphics[scale=.48,angle=-90]{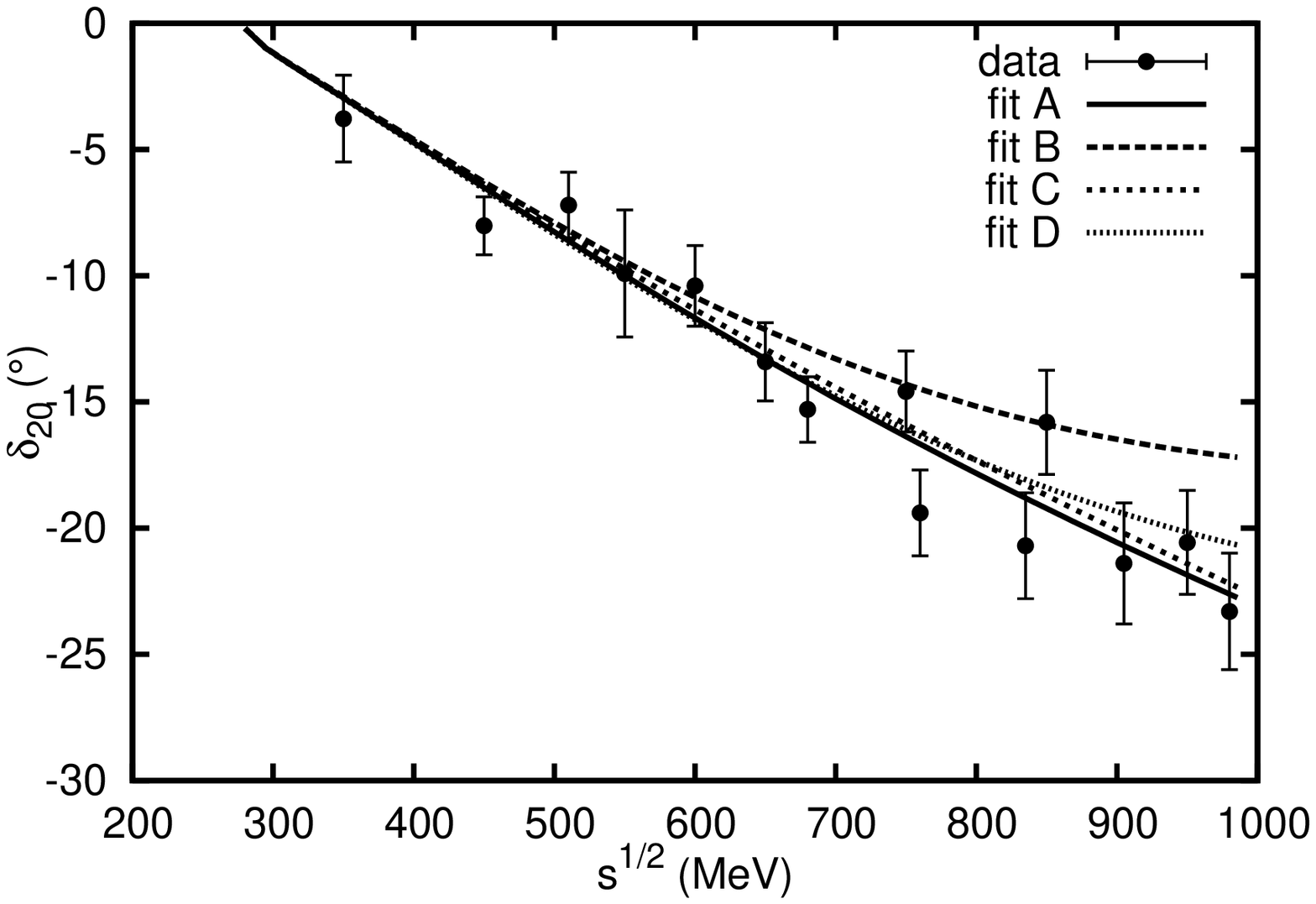}     
      \includegraphics[scale=.48,angle=-90]{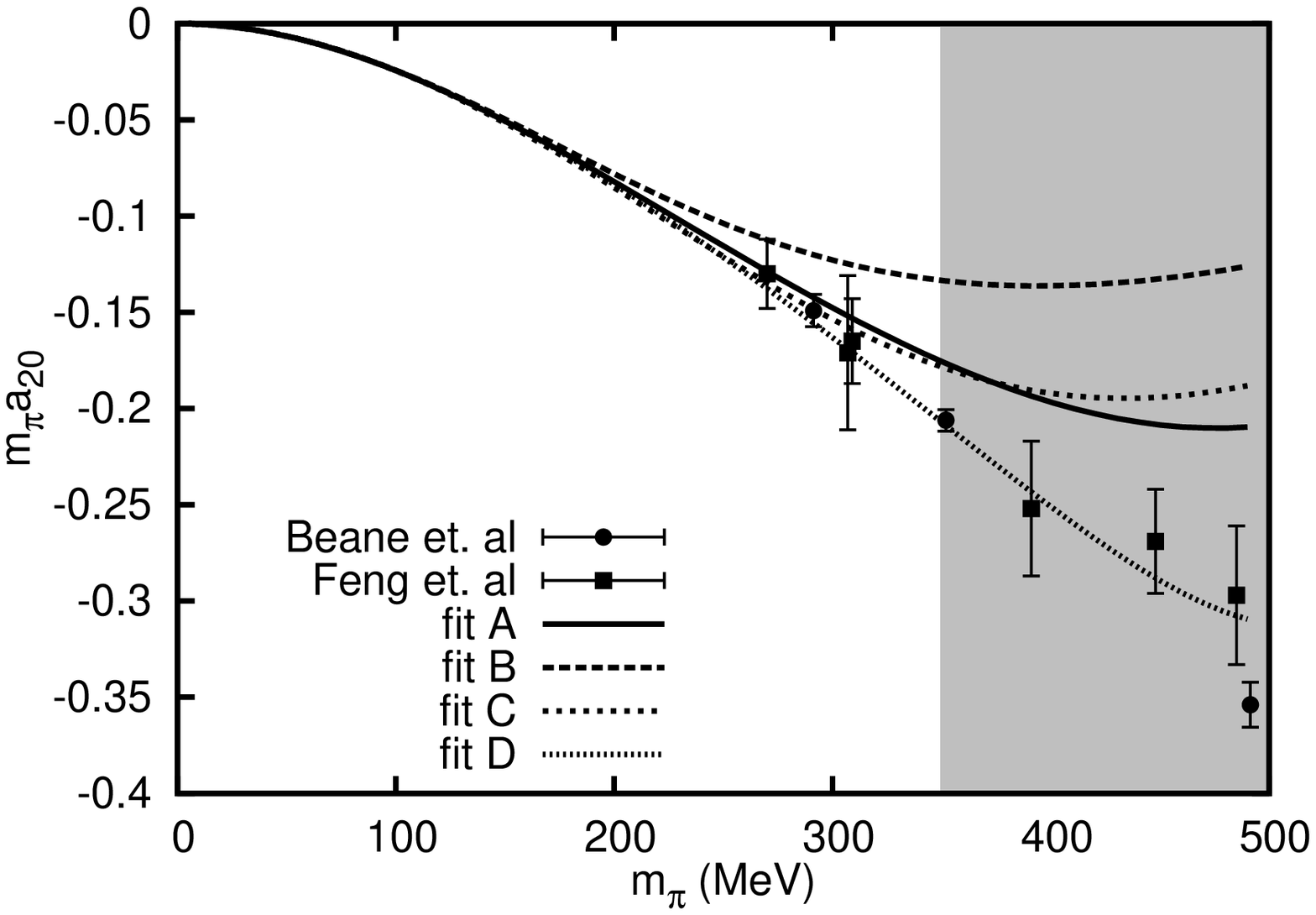}
    }
  }
  \caption{Different fits to physical $\pi\pi$ scattering phase shifts $\delta_{IJ}$
and lattice results on the $\rho$ mass, $f_\pi$ and the $a_{20}$ scattering length.
For the lattice results, we have not extended our fits to the grey area ($m_\pi>350\,$MeV, although 
it is displayed  to show how the fits deteriorate, or not, beyond 350 MeV.
The phase shift data comes from \cite{data,Rosselet:1976pu,Grayer:1974cr} and the 
lattice results from \cite{lattice1,lattice2,lattice3,lattice4,
lattice5,fpilattJLQCD,Beane:2007xs,Feng:2009ij}.
 }
  \label{fita}
\end{figure*}

We have thus fitted the mIAM to the elastic 
pion-pion scattering phase shifts shown on the left column
of Fig.\ref{fita}. Let us remark that we have fitted data up to 1 GeV
for the $(1,1)$ and $(2,0)$ waves and up to 800 MeV
for the $(0,0)$ channel; beyond that energy the effects of
the  $f_0(980)$ are important and 
cannot be reproduced with our single channel formalism.
There are, of course,  coupled channel unitarization
formalisms \cite{Oller:1997ti,IAMcoupled}, which are very successful and generate the $f_0(980)$ among other resonances, but they
lie beyond the scope of this work, 
mostly because of their simpler treatment of the left cut -- if dealing with it at all--,
or their dependence on additional parameters, all of which can introduce further model dependences.

In addition, in order to constrain further
the LECs that govern the pion mass dependence, we have also fitted
the lattice results shown in Figs.1 and 2 
in previous sections, although only up to a pion mass of 350 MeV.  

Still, we have 11 parameters to fit, and 
even with these constraints, because of the large correlations
between parameters, during the fitting procedure some LECs can take
values very far from their typical ones, even of 
different order of magnitude, for tiny improvements in
the data $\chi^2$. For that reason, we have also considered an averaged
$\chi^2_{LECs}$ term as a constraint for our fits, 
that measures how far the LECs are from 
some reference values that we provide in Table~\ref{LECs}.
To further constrain the fit, we will also require
the $\rho$ mass and width leading $1/N_c$ scaling to follow a
$\bar qq$ pattern, so we also consider  
a $\chi^2_{\rho-\bar qq}$ measure, as described in \cite{Pelaez:2006nj},
to constrain the $\rho$ behavior to that of a $\bar qq$. 
Note that uncertainties for $N_c=3$ are of the order of 30\%,
but this constraint becomes stricter as $N_c$ grows. Nevertheless,
we will 
never apply this constraint for $N_c$ larger than 20,
since otherwise the theory would become weakly interacting and the whole
unitarization procedure would loose sense, as repeatedly 
explained in \cite{Pelaez:2010er,Pelaez:2009eu}.

\begin{figure}
  \centering
  \includegraphics[scale=.48,angle=-90]{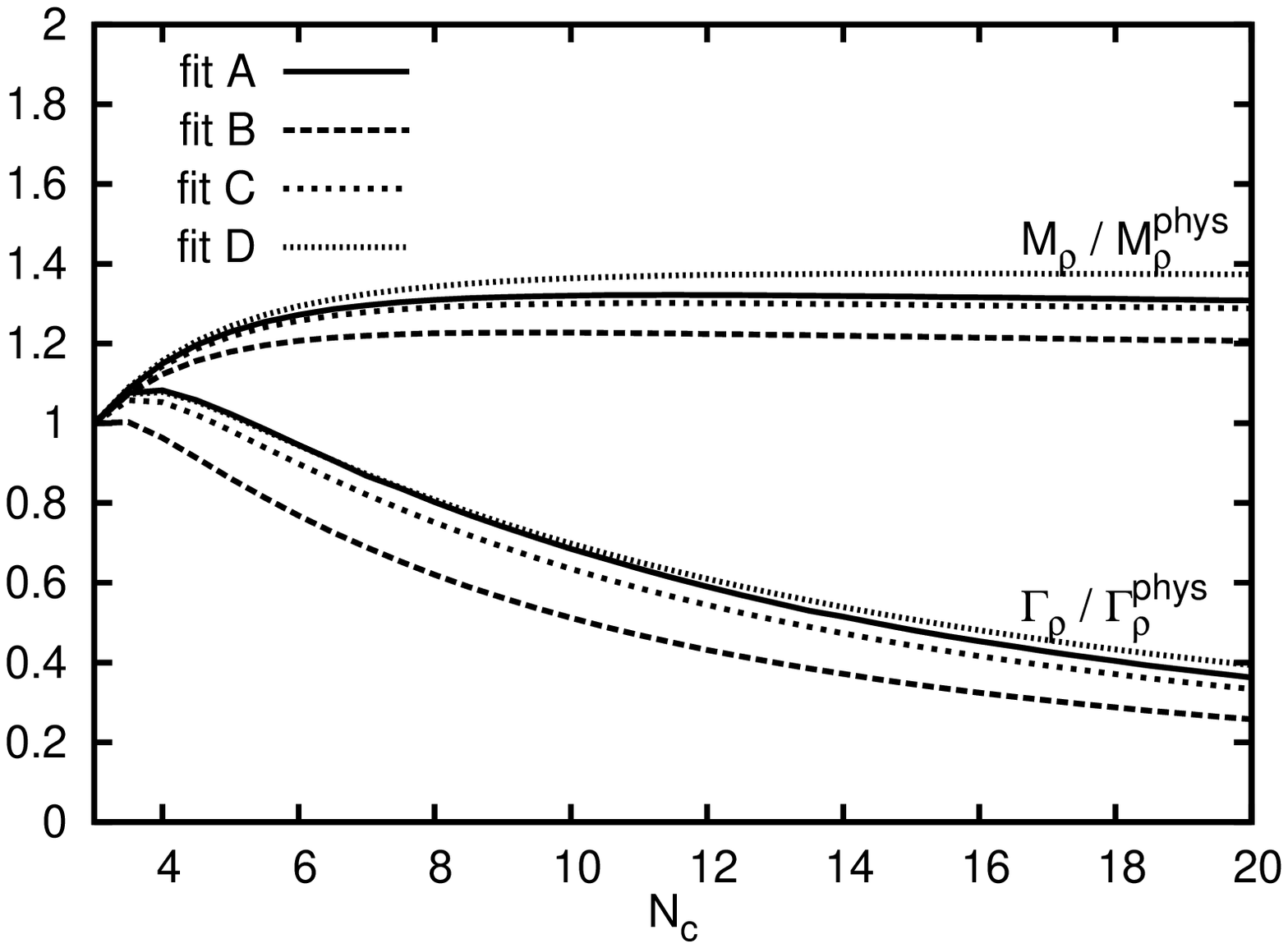}
  \caption{$N_c$ dependence of $\rho(770)$ mass and width for the different fits
described in the text.}
  \label{rhoNc}
\end{figure}

In summary, we are considering several $\chi^2$-like
functions that, when smaller or close to 1, ensure
a good description of each feature described above:
$\chi^2_{data}$,  $\chi^{2\,lattice}_{M_\rho}$, $\chi^{2\,lattice}_{a_{20}}$,
$\chi^{2\,lattice}_{f_\pi}$
$\chi^2_{\rho-\bar qq}$ and $\chi^2_{LECs}$.
The problem is that many of these are not really well defined $\chi^2$ functions
in the statistical sense. The reasons are that
some uncertainties we use are theoretical (as for the $1/N_c$ behavior), 
and that some sets, both for real data or lattice,
 are incompatible with each other and we have to guess some systematic uncertainty of the order of the difference between different sets.
In addition, the number of ``data points''
to be fitted for each feature is very different, and we could get a bad 
description of one feature with few ``data'' at the expense of a tiny improvement on another feature with more ``data'', but affected by systematic errors
crudely estimated.
Thus, there is not a single fit of data and lattice
minimizing the sum of all $\chi^2$ functions, since we do not know how to weight each one of them against the others.
We are nevertheless presenting four different ``fits'' A, B, C and D 
where we have imposed that 
each one of the $\chi^2$ should be relatively close or smaller than 1. 
This is still quite a strong constraint,
and ensures, as we will see in Fig.~\ref{fita}, 
that all features are fairly well described up to the applicability region. 
Thus, in Table~\ref{LECs} we show the parameters of 
four different fits, 
which, as seen in Fig.\ref{fita}, cover the 
data and lattice results fitted up to pion masses of 350 MeV.
In particular, we show the $M_\rho$ dependence on $m_\pi$ versus the
lattice data, the $IJ=00,11,20$ partial waves,  as well as the $a_{20}$ and
$f_\pi$ dependence on $m_\pi$ compared with lattice results.
Note that these fits tend to prefer the
 somewhat stronger $M_\rho$ pion mass dependence
found by the MILC \cite{lattice1}, ETMC \cite{lattice2} and RBC-UKQCD \cite{lattice3}
collaborations.
The leading $1/N_c$ behavior of the $\rho$ pole is also 
shown in Fig.~\ref{rhoNc}.

\begin{table}[h]
  \centering
  \begin{tabular}{|c|cccc|c|}
    \hline
    Fit         & Fit A &Fit B   & Fit C  & Fit D  & Reference  \\
    \hline             
    $10^3l_1^r$ & -5.0 &  -4.7   & -5.0  & -4.0  &$-3.3\pm2.0$ \\
    $10^3l_2^r$ & 1.7  &  0.95   &  1.7  &  1.24  &$1.9\pm1.0$ \\
    $10^3l_3^r$ & 0.82 &  0.82   &  -6.0 &  0.82 &$0.82\pm3.8$   \\
    $10^3l_4^r$ &  6.5 &  4.96   &  3.5  &  6.5  &$6.2\pm2.0$   \\
    \hline                      
    $10^4r_1$   & -0.6 &   -1.0  & -0.7  & -0.6  & -0.6   \\
    $10^4r_2$   &  1.3 &   1.3   & 3.7    & 1.5   & 1.3   \\
    $10^4r_3$   & -1.7 &  -0.29   & 2.7   & -3.3   &-1.7   \\
    $10^4r_4$   & 2.0  &   4.2   & 2.8   & 0.95   &-1.0   \\
    $10^4r_5$   & 2.0  &   2.3   & 2.0   & 1.7   &1.5   \\
    $10^4r_6$   & -0.56&  -0.98  &-0.5   &-0.7   & 0.4  \\
    \hline
    $\tilde r_f$ & -3.4 & -1.8 & -2.3 & -4.6 & 0   \\
    \hline
  \end{tabular} 
  \caption{LECs for the $O(p^6)$ fits A, B, C and D and 
the reference values we use in $\chi^2_{LECs}$.
For the $O(p^4)$ LECs we have used values that cover the different sets
in Table I, whereas for the $O(p^6)$ LECs we show 
estimates from 
\cite{Bijnens:1997vq,Colangelo:2001df}
}
  \label{LECs}
\end{table}

In view of the
different $\chi^2$ per data points (abbreviated as $\chi^2/{\rm (\# points)}$)
that we list in Table~\ref{tab:chi2} and will describe in detail below,
we consider that fits A and D are the best ones, 
particularly because their LECs are quite compatible with
the values of the $O(p^4)$ LECs and reasonably close to the crude expectations for the
$O(p^6)$ ones. 
The difference between them is that fit A 
describes better the $M_\rho$ pion mass dependence --  even up to 
very high pion masses -- and not so well the $a_{20}$ scattering length, 
whereas fit D does the opposite.
In addition, we provide two other fits to illustrate the uncertainties.
First, on fit B  we try to force the softer
 $M_\rho$ pion mass dependence found by the CP-PACS-Adelaide 
\cite{lattice4}
and QCFSF \cite{lattice5} collaborations.
Note that this fit then prefers the less 
negative phase shift data of the $(I,J)=(2,0)$ channel, 
an, as a consequence it does not describe 
very well the mass dependence of the  $a_{20}$ scattering length.
 Also, the agreement with the LECs estimates is worse than for fit A, as seen in Table~\ref{LECs}. 
In addition we are showing the results for fit C, which is very similar to A
and D
up to pion masses of 300 MeV, but shows a rather 
dramatic drop of the $\rho$ mass above 350 MeV. It is a general 
feature of our fits that they can be made reasonably compatible
 with data up to 350 MeV, but beyond that point they can start
 diverging widely for not very large changes in the LECs. That 
is the reason why we consider that our approach is 
not reliable beyond a pion mass of 300 to 350 MeV, depending on the observable. 

Before proceeding to the next section we want to provide, 
for each feature included in our fit, the $\chi^2$
per number of data points. Of course, in view of the
clear incompatibilities of different sets of data for a given quantity
 -- from experiment or lattice --, we 
have either added some systematic uncertainty
 to cover different sets, or fitted
some particular subsets of data, as we detail next.
 Let us emphasize that, since we use the IAM and 
have made several approximations on its derivation, 
we are obviously not aiming at precision. As it is seen in Fig.\ref{fita}
the spread of our four fits covers roughly the different data points.

\begin{table}
  \centering
  \begin{tabular}{|c|cccc|}
    \hline
    $\chi^2/({\rm \# points})$ & Fit A & Fit B & Fit C & Fit D \\\hline
    $\delta_{00}$ Sol. B \cite{Grayer:1974cr} & 0.8 & 1.1  & 0.5  & 0.6  \\
    $\delta_{00}$ Sol. C \cite{Grayer:1974cr} & 5.6 & 1.1  & 2.6  & 5.7 \\
    $\delta_{00}$ below 400 MeV & 0.87  & 0.8 & 0.9  & 1.0 \\
    $\delta_{11}$  & 1.2 & 1.6 & 0.7  & 0.7 \\
    $\delta_{20}$  & 0.2 & 1.0  & 0.3  & 0.4 \\
    $M_\rho^{lattice}$ & 0.77 & 0.68  & 0.88  & 1.35 \\
    $a_{20}^{lattice}$ & 0.2  & 4.0  & 0.06  & 0.2 \\
    $f_\pi^{lattice}$  & 1.4 & 1.1  & 0.6  & 0.8 \\
    $N_c$ & 1.4 & 0.6  & 1.2  & 1.5 \\
    LECs  & 1.5   & 3.8  & 3.2  & 1.4 \\
    \hline
  \end{tabular}
  \caption{$\chi^2/({\rm \# points})$ for each feature fitted. See
  main text for details of the $\chi^2$ calculation.}
  \label{tab:chi2}
\end{table}

The chaotic situation with the (0,0) phase shifts
above 450 MeV (obtained from $\pi N\rightarrow \pi\pi N$ experiments)
is well known and very visible in Fig.\ref{fita}. Even the same CERN/Munich 
experiment \cite{Grayer:1974cr}
provides 5 different analysis (A, B, C, D and E)
incompatible with each other. Among them, data sets B and C have been shown
to be the ones that satisfy better several dispersive constraints \cite{Pelaez:2004vs}. Thus, we have added
linearly 2 degrees of systematic uncertainty  
to the points of these two sets -- not much given the 
huge incompatibilities. In Table \ref{tab:chi2} we provide the resulting 
$\chi^2/{\rm (\# points)}$ for each fit with respect to
the data sets B and C in \cite{Grayer:1974cr}. Note that, for each fit,
at least one of the two experimental sets is well described, i.e., with
 $\chi^2/{\rm (\# points)}<1$.
If we wanted $\chi^2/{\rm (\# points)}\simeq1$ for all data sets 
plotted in Fig.\ref{fita} simultaneously, 
we would have to add 5 degrees as a 
systematic error, which again would not be  much taking into account that differences 
between data sets in the 700 to 800 MeV region can be as high as 15 degrees.
Fortunately, the $K_{l4}$ decay data below 400 MeV \cite{Rosselet:1976pu}
is of much better quality,
but is easily reproduced by our fits
without the need to add systematic uncertainties 
as seen also in Table~\ref{tab:chi2}.

The fits to other quantities are less complicated. Despite they look relatively
close, the two sets of $(I,J)=(1,1)$ $\pi\pi$ scattering phase shifts
are not compatible, since they provide tiny statistical errors only.
Their difference is of the order of two degrees, 
which once again we have added linearly as a systematic error.
Concerning the $(2,0)$ wave, once more it is 
clear that there are incompatible sets of data, and we have added 
the same  2 degrees of systematic uncertainty.
The resulting $\chi^2/{\rm (\# points)}$ for all our fits
are given in Table~\ref{tab:chi2}.
Note that, once again, fit B can be used  
to differentiate the $(2,0)$ ``low data'' set
from the ``high data'' set and that it does not reproduce so well as the others the physical
$\rho$ shape.

Concerning the lattice predictions for the $\rho$ mass, 
there is also an obvious conflict between different collaborations,
with differences in the $M_\rho$ value as high as 70 MeV. Hence, the
$\chi^2/({\rm \# points})$ given in Table~\ref{tab:chi2} has
been calculated for the points with $m_\pi$ below 350 MeV,
adding a systematic uncertainty of 35 MeV.
Note that if we took into account data up to
 $m_\pi=400\,$MeV, the $\chi^2/({\rm \# points})$
would have been 0.95, 0.93, 0.99 and 2.1 for fits A to D, respectively.

In the case of the $f_\pi$ lattice results, different collaborations 
have points
clustered sufficiently  close in groups of three  to estimate 
a systematic uncertainty as half the difference between
the highest and lowest data point, 
which is added linearly to the statistical errors.
The resulting $\chi^2$ for each fit with this prescription is shown in
Table~\ref{tab:chi2}. Note that we
find some difficulty in describing the JLQCD results simultaneously with the other features.

The leading order $1/N_c$ behavior of the $\rho$ has been adjusted to be 
exactly that of a pure $q\bar q$ state. This is believed to be 
a very good approximation, although given its large physical 
width it may easily have some small $\pi\pi$ component, that we neglect.
Using the $\chi^2$ definitions and estimating
the uncertainty to be exactly $1/N_c$ times the leading term, as explained in
\cite{Pelaez:2006nj}, we find the 
averaged $\chi^2$ for the $\rho$ as a pure 
$q\bar q$ state shown in Table \ref{tab:chi2}.

The last quantity that we have fitted is the (2,0) 
scattering length, where we find
one data point right at $m_\pi=350\,$ MeV and four other points below.
There are two collaborations and their points are 
relatively consistent with each other. 
Thus, and despite the large differences 
between different collaborations for other observables, 
we have not added any systematic
error to these points. For the four points strictly below 350 MeV we 
show the $\chi^2/{\rm (\# points)}$ for each fit in table \ref{tab:chi2}.
Once again fit B has serious problems describing this observable, but
note that it
describes better the ``high data'' set of
(2,0) phase shifts. If we now include the point at $m_\pi=350\,$MeV 
with its tiny 
uncertainty, the $\chi^2/{\rm (\# points)}$ of all fits except
fit D grow beyond 4.5.

It is clear that our fits A, C and D give a very 
good general description of, at least,
one set of data for each observable up to 300 MeV, 
but not so good up to 350 MeV,
where they start deviating from each other 
also for the $M_\rho$ quark mass dependence.
For that reason we will consider our 
IAM approach to be valid only up to, roughly, the 
300 to 350 MeV region. This is, of course a crude estimate and varies from 
one observable to another.

Finally, concerning the $\chi^2_{LECs}$, let us remark that to calculate it
we use the reference values given in Table \ref{LECs}. 
Since the values of the $r_i$
$O(p^6)$ parameters come from resonance saturation estimates 
(with resonances of angular momentum smaller than 2), 
we have assumed that they 
are only correct in the order of magnitude
and therefore have a 100\% uncertainty. 
For all fits, most of the deviations come from 
the $O(p^6)$ LECs, which are the worst known.
Note that, in general, but particularly in this case, 
fits A and D are the ones with
 better $\chi^2$ and that is the reason why we consider them as our best fits. 
Fits B and C are just 
given for illustration of different scenarios and to show that
the {\it predictions} that we will detail next are robust 
even allowing for larger uncertainties or deviations from our best fits.

\subsection{Predictions from the fits}
\label{sec:predictions}

Once we have obtained a relatively good description of the data and 
existing lattice results on certain observables, we will now use these three 
fits to obtain {\it predictions for other observables}. 
In general, the spread between our curves should be 
considered as a naive indication of our systematic uncertainties. 

\subsubsection{The $\rho$ width and coupling}

\begin{figure}
  \centering
\vbox{
      \includegraphics[scale=.5,angle=-90]{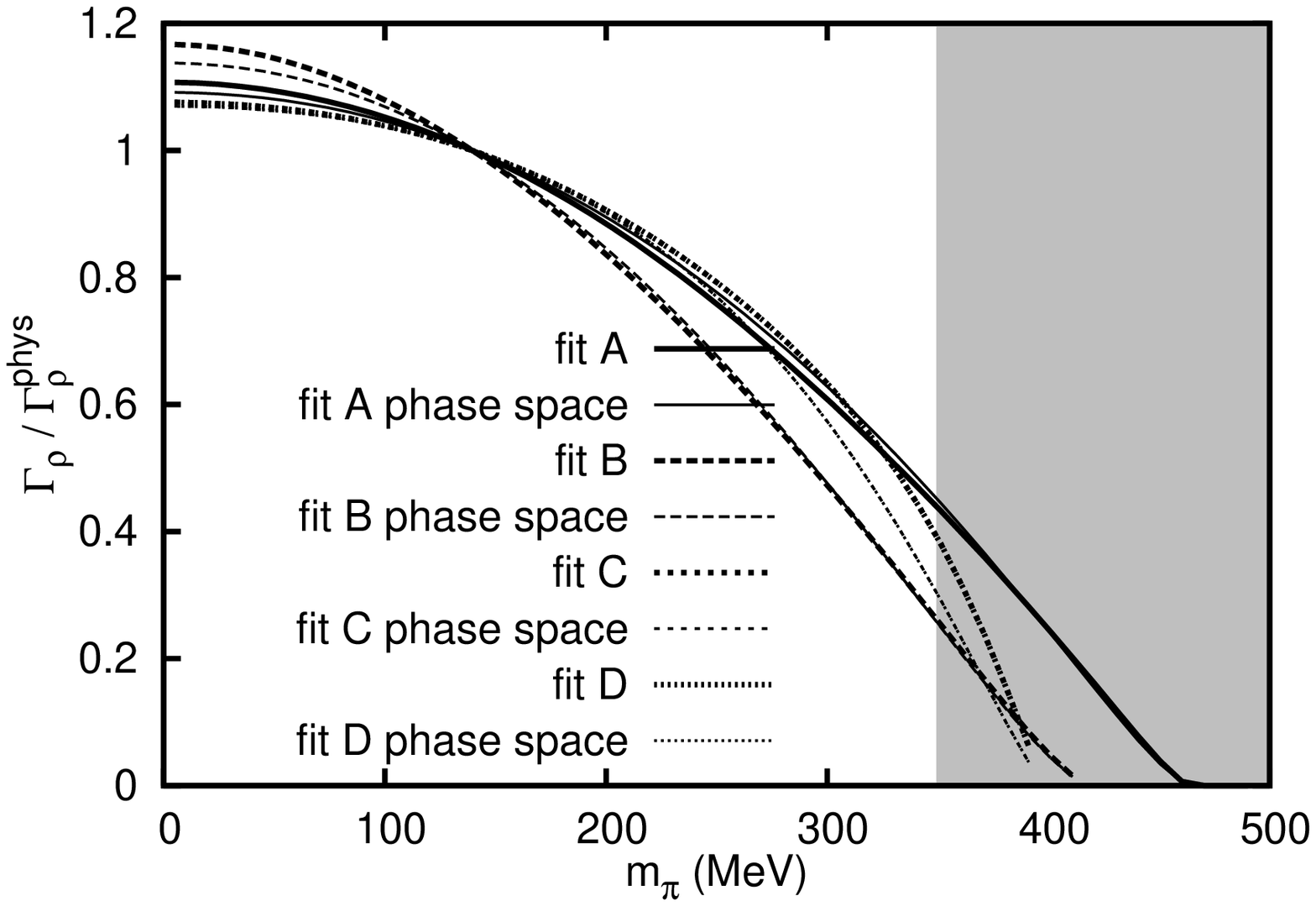} 
      \includegraphics[scale=.5,angle=-90]{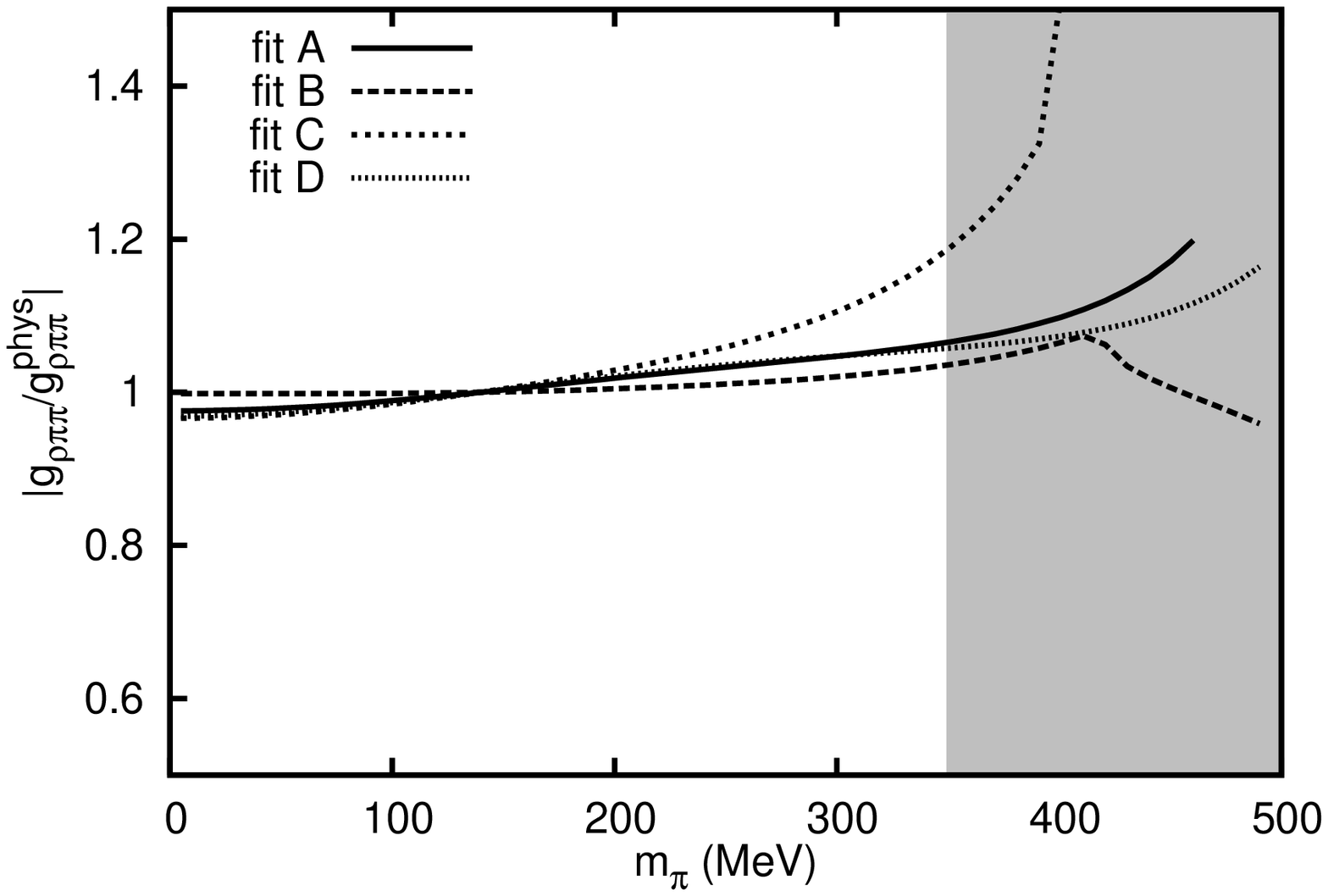}     
      \includegraphics[scale=.5,angle=-90]{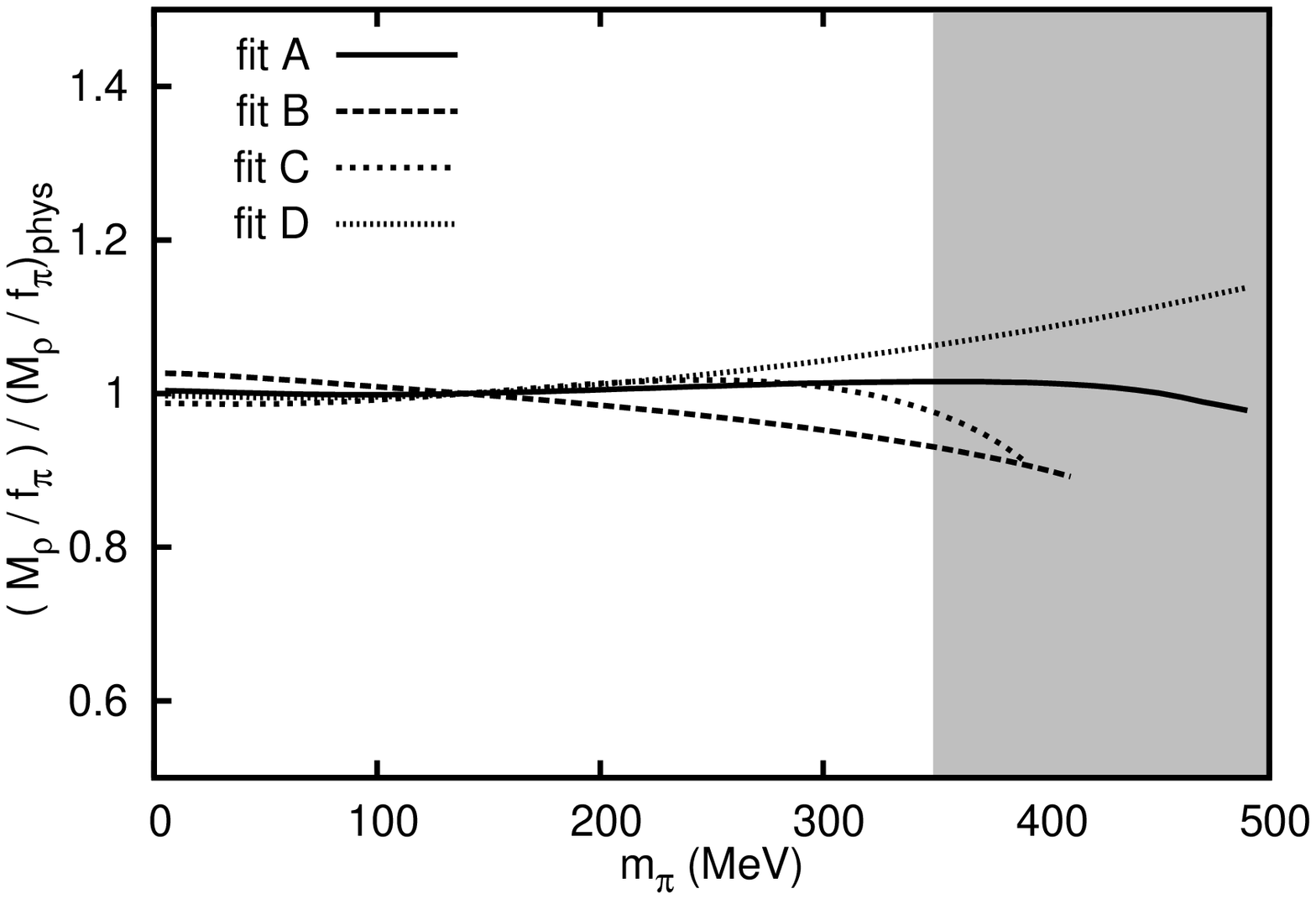}  
}
  \caption{Predictions for the pion mass dependence of $\rho$ parameters 
from the IAM two-loop fits
 described in the text and Fig.\ref{fita}. Top: We show the rho width 
from each IAM fit together with the
behavior expected only from phase-space variation 
due to the changing $\rho$ mass. 
Middle: the $\rho$ coupling. Bottom: the $M_\rho/f_\pi$ ratio. 
The last two are remarkably independent 
of the quark or pion mass, which ensures that 
the KSRF relation is not spoiled within the applicability region of our approach. }
\label{rhofigpredictions}
\end{figure}

\begin{table}
  \centering
  \begin{tabular}{|c|c|c|}
\hline
 & $\sqrt{s_\rho}$ (MeV) & $\vert g_\rho\vert$ \\ \hline
fit A & $754-i\,74$ & 6.1 \\ \hline
fit B & $772-i\,71$ & 5.9 \\ \hline
fit C & $759-i\,72$ & 6.0 \\\hline
fit D & $763-i\,71$ & 5.9 \\
\hline
  \end{tabular}
  \caption{Values of the $\rho$ pole position in the lower 
half plane of the second Riemann sheet and $\rho\pi\pi$ couplings for the different fits described in the text.}
\label{tab:rhopolecoupling}
\end{table}
First of all, in Table~\ref{tab:rhopolecoupling} we provide the values of the $\rho$ pole position, $s_\rho$,
in the lower half plane of the second Riemann sheet and its coupling to two pions 
defined from the (1,1) partial wave as 
\begin{equation}
g_\rho^2=-16 \pi\lim_{s \to s_{\rm pole}}
(s-s_\rho)\ t_{11}(s)\frac{3}{4\, p^{2}}\label{eq:residue},
\end{equation}
where the normalization factors are chosen 
to recover
the usual expression for the two-meson
width of the $\rho$:
\begin{equation}
\Gamma_\rho=\vert g_\rho \vert^2\frac{1}{6\pi}\frac{|{\bf p}|^3}{M_\rho^2}.
\label{eq:V-width}
\end{equation}
If we approximate $M_\rho={\rm Re}\,\sqrt{s_\rho}$ in the above equations,
and we use the value of the coupling in Table \ref{tab:rhopolecoupling},
we find that the width of the $\rho$ for fits A to D are 149, 144, 145
and 142 MeV, respectively, to be compared with the PDG value of $149\pm1\,$MeV.
\cite{PDG}.
A rather good description, given the approximation, the quality of the data,
and that we only provide central values.

On the top panel of Fig.\ref{rhofigpredictions} 
we show the evolution of the
$\rho$ width for 
the two-loop unitarized ChPT fits described in the text and Fig.\ref{fita}.
Although hard to see because they are almost overlapping, 
we provide, together with the result of each fit, the expected 
variation from phase space  due  to the change in $M_\rho$ only,
assuming a constant $\rho-2\pi$ coupling
in a Breit-Wigner form. As it happened in the one-loop case 
\cite{chiralexIAM}, we see that
this constancy assumption of the $\rho$ coupling is a very good approximation, 
so that this feature is rather robust under higher order ChPT corrections
when fitting to data and lattice results on $f_\pi$ and $a_{20}$.

Moreover, in the middle panel of Fig.\ref{rhofigpredictions}, we show the
actual two-loop IAM calculation of the coupling from the residues of the amplitude at the resonance pole. 
We can see that it is rather independent of 
the pion mass up to $m_\pi\simeq 350$ MeV. Namely,
fit B  barely changes at all, fits A and D 
change by roughly 5\%  and only fit C changes by 15\%, to be compared with the
 factor of 2.5 times that the pion 
mass is increased from its physical value up to 350 MeV. 
Actually, this corresponds to
an increase in the quark mass larger than a factor of 6.
Thus, the relatively weak dependence of 
the $\rho-2\pi$ coupling on the quark mass
is confirmed at two loops. This is of particular 
relevance to some lattice calculations
that have assumed a constant coupling in studies 
of the $\rho$ width \cite{Aoki:2007rd}. 
Nevertheless, this prediction of $M_\rho$ with the
quark mass is limited to pion masses below roughly 350 MeV, and 
more uncertain within our two-loop approach than at one-loop,
as we see from the spread between different fits. 
Actually, for the less favored fit C we see up to a 15\% variation, 
but note that the $M_\rho$ mass
in this fit C starts behaving rather weird around 350 MeV, 
and most likely it is, very roughly speaking, only reliable up to the region between
300 to 350 MeV, depending on the observable.

Finally for the $\rho(770)$,  we show in the 
lower  panel of Fig.\ref{rhofigpredictions} 
the evolution of the $M_\rho/f_\pi$ ratio, which comes
 almost independent of $m_\pi$ in all of our two-loop fits.
This ratio is not really a prediction, since we have fitted both $M_\rho$ and $f_\pi$
to lattice data.
However, it is of particular interest 
for the well known KSRF relation \cite{KSRF}, which 
provides an striking connection between 
the $\rho-2\pi$ coupling and the $M_\rho/f_\pi$ ratio:
\begin{equation}
g_{\rho\pi\pi}^2\simeq M_\rho^2/8f_\pi^2,
\end{equation}
and holds fairly well for the physical values of these constants.
Since we have just checked the almost constancy of $g_{\rho\pi\pi}$, 
then, an almost constant $M_\rho/f_\pi$ 
ratio as found in  Fig.\ref{rhofigpredictions} means that the KSRF relation
also holds rather nicely, at least within our applicability region. 
This corroborates the one-loop
results already found in unitarized SU(3) ChPT \cite{Nebreda:2010wv}.

\subsubsection{The $\sigma$ parameters}

Let us turn to the predictions for the pion mass dependence of the 
$\sigma$ or $f_0(600)$ scalar resonance, whose nature is still 
very controversial. As we did with the $\rho$ we provide first the
values of the resulting $\sigma$ pole positions and couplings 
in Table \ref{tab:sigmapolecoupling}, obtained as follows:
\begin{equation}
g^2=-16 \pi \lim_{s \to s_{\sigma}}
(s-s_{\sigma})\ t_{00}(s)\label{eq:residuescalar}.
\end{equation}

For comparison we also provide in the last rows of 
Table~\ref{tab:sigmapolecoupling} results from other works in the literature.
Let us remark that, without using the 
recent and very precise $K_{l4}$ decay data \cite{Rosselet:1976pu},  
the $\sigma$ pole
was obtained more than fourteen years ago 
at $\sqrt{s_\sigma}=440-i\,245\,$MeV, with the single channel 
one loop IAM in \cite{Dobado:1996ps}. 
\begin{table}
  \centering
  \begin{tabular}{|c|c|c|}
\hline
 & $\sqrt{s_\sigma}$ (MeV) & $\vert g_\sigma\vert$ (MeV) \\ \hline
fit A & $453-i\,265$ & 3.4 \\ \hline
fit B & $474-i\,248$ & 3.5 \\ \hline
fit C & $466-i\,245$ & 3.3 \\ \hline
fit D & $453-i\,271$ & 3.5 \\ \hline\hline
H. Leutwyler et al.\cite{Leutwyler:2008xd}& $441^{+16}_{-8}-i (272^{+9}_{-12.5})$&$3.31^{+0.35}_{-0.15}$\\ \hline
R. Garc\'{\i}a-Mart\'{\i}n {\it et al.} \cite{Yndurain:2007qm}& $474\pm6-i (254\pm4)$& $3.58\pm0.03$\\ \hline
R. Kaminski {\it et al.}\cite{Kaminski:2009qg}& $442-i 290$& $2.47\pm0.45$\\ \hline 
J.A. Oller \cite{Oller:2003vf}& $(443\pm2)-i(216\pm4)$ & $2.97\pm0.04 $\\
\hline
  \end{tabular}
  \caption{Values of the $\sigma$ pole position in the lower 
    half plane of the second Riemann sheet and $\sigma\pi\pi$ 
    couplings for the different fits described in the text and some references
in the literature.}
\label{tab:sigmapolecoupling}
\end{table}

The $O(p^6)$ results
are in quantitative agreement with the $O(p^4)$ ones for pion masses
 lower than about 300 MeV.
For instance, as the
quark mass is increased, the 
relative growth of the $\sigma$ mass, defined as the
real part of the $\sigma$ pole position, is slower
than the pion mass growth, but still somewhat faster than for the $\rho$ mass.

\begin{figure}[h]
  \centering
  \vbox{
  \includegraphics[scale=.48,angle=-90]{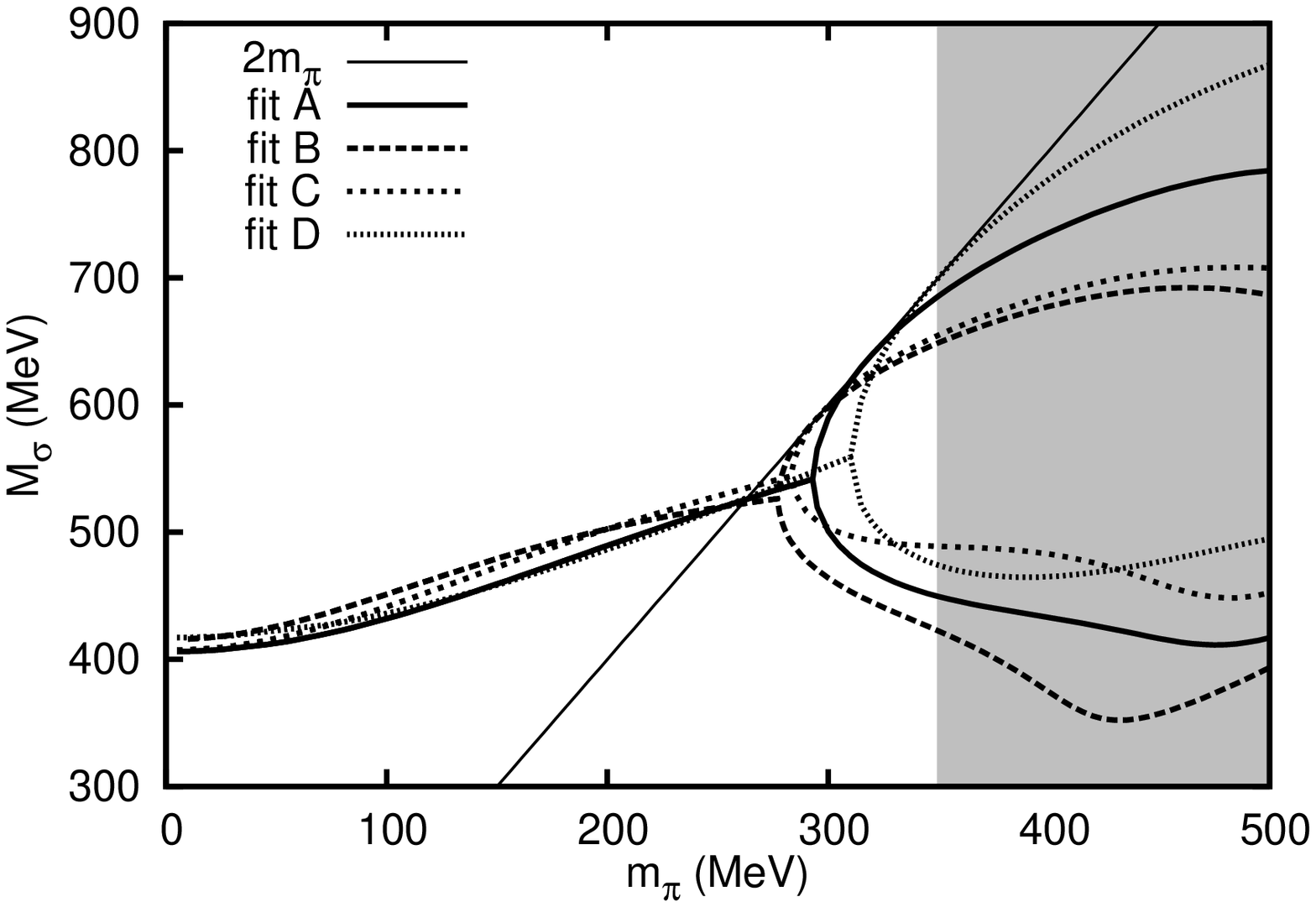}
  \includegraphics[scale=.48,angle=-90]{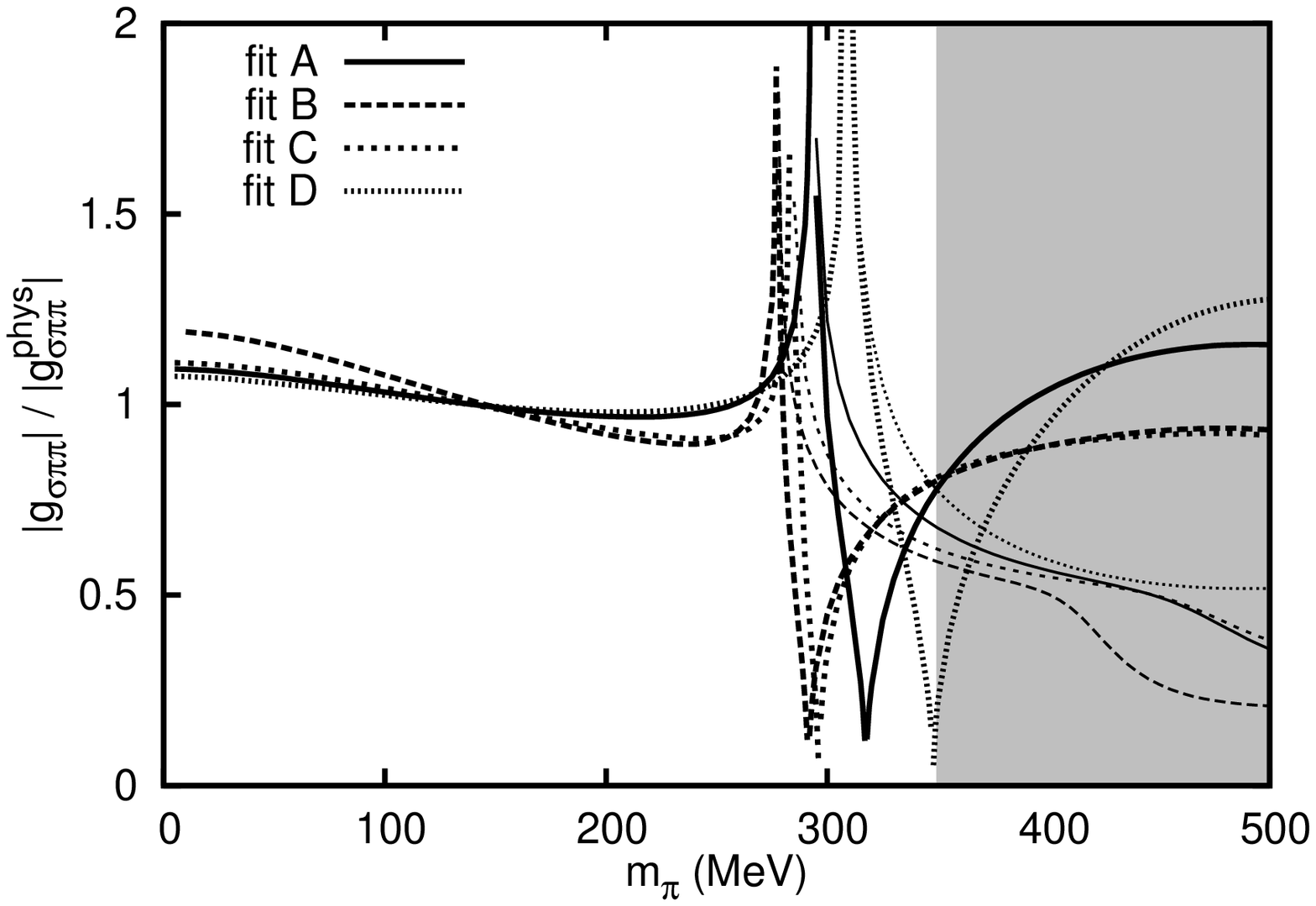}
}
  \caption{Predictions for the $\sigma$ from the IAM two-loop fits described in the text and Fig.\ref{fita}.
Note they are remarkably stable despite the differences between fits 
for the $\rho$. We show as a grey area the region where $m_\pi>350\,$ MeV, which marks the applicability bound of the approach. Top: we show 
the predicted pion mass dependence for the $\sigma$ pole mass. Note the appearance of two branches around 300 MeV, as explained in the text. Bottom:
we show the dependence of the $\sigma$ 
coupling to two pions, which, as explained in the text, is rather strong
particularly as the pion mass approaches the 
value where the 
two conjugate poles on the second Riemann sheet reach the real axis. 
For each fit, the thick line corresponds to the upper branch and the thin one to the lower branch.
 }
\label{sigmafigpredictions}
\end{figure}

As we saw for the one-loop case in previous sections and in \cite{chiralexIAM},
as the pion mass grows, since the sigma mass grows slower, its width
becomes narrower and narrower, and 
its two conjugate poles approach the real axis. 
But contrary to the $\rho$ case, it is possible to
have poles below threshold on the second Riemann sheet, 
and the $\sigma$ conjugate
poles actually meet at some point on the real axis below threshold. 
If the pion mass keeps on increasing,
 both poles stay on the real axis, but one moves very little,
remaining at masses lower than threshold, 
whereas the other one increases its mass, until it eventually jumps to the first Riemann sheet. 
All these features occur once again at two loops and this
double real pole structure is nicely seen
in the top panel of Fig.\ref{sigmafigpredictions} 
as a double branch for each fit.

 Let us remark that the appearance of two branches 
is not an artifact of the IAM, but is a general feature of 
scattering theory of scalar amplitudes with poles 
close to threshold \cite{inprep}, also seen in other contexts \cite{othercontexts}.
It is just the way scalar poles approach to threshold as one
 changes the features of the interaction. Namely, there are no restrictions on 
where a scalar pole should be on the real axis 
below threshold {\it on the second Riemann sheet}, except that
poles appear in conjugate pairs out of the real axis, 
or on the real axis below threshold. 
In the first case, they obviously have the same ``mass'', but 
if this pair reaches the real axis, the two poles
no longer have to be conjugated and hence the double branch is a general feature.
In contrast, all non-scalar waves have centrifugal $p^{2J}$ factors 
relevant around threshold, 
that force their  second sheet poles to reach the 
real axis precisely at threshold \cite{inprep} where one of them jumps into the first
sheet whereas the other stays in the second, as it happens here with the $\rho(770)$.

The IAM, of course, not only reproduces this general feature,
but also provides an estimate of the pion mass where this apparent splitting occurs,
which is not generic, but a specific value due to the QCD dynamics
underlying the properties of the lightest scalar meson.

Let us remark that all our fits are very
consistent with each other 
for the $\sigma$, despite their differences for the $\rho$ behavior. 
As explained above, this is due to the fact that in the generation of the 
$\sigma$ 
the chiral loops play a dominant role. At NLO these are independent of 
the LECs, and at NNLO they only depend on the $O(p^4)$ LECs,  but not on the $r_i$.
Therefore the $\sigma$ avoids most of the largest uncertainties that affect the $\rho$
and, as a consequence, the results for the $\sigma$ are much more robust.
There are however, some quantitative differences with the one-loop results
in \cite{chiralexIAM}: for instance, the point where the two conjugate poles of the
$\sigma$  meet on the real axis
 occurs for $m_\pi$ masses about 20 MeV lower; namely, at $m_\pi= 280-310\,$ MeV for the different
two-loop fits versus $300-330\,$MeV for the one-loop description. This is a rather small
correction to the one-loop result and confirms the robustness of our results for the $\sigma$, 
even under higher loop corrections, at least up to $m_\pi\simeq 300-350$ MeV,
depending on the observable. 
However, for higher pion masses the quantitative spread is much larger,
although
the four fits yield the same qualitative 
predictions for the poles in the two branches. 
Closely related to the decrease of the ``splitting point'' is the fact that 
the pole of the "upper branch" reaches the threshold faster 
than in the one-loop case as the pion mass grows. Note that the threshold variation 
corresponds to the line labeled $2m_\pi$ 
in the figure, that the upper branches of all fits touch 
very soon after the two branch splitting, around $m_\pi\simeq290-350\,$MeV, versus $m_\pi\simeq460\,$ MeV
for the one loop calculation in \cite{chiralexIAM}. 

The relevance of these results is that, when the upper branch pole reaches threshold, it jumps into the first 
Riemann sheet, and becomes a usual bound state. (One might wonder if the dispersion relation for the IAM applies now that there is a pole on the first Riemann sheet, but note that the IAM derivation is obtained from dispersion relations for the {\it inverse amplitude}, so that this pole on the first Riemann sheet
is a zero for the inverse amplitude and therefore does not alter the analytic structure).
Our two-loop results seem to indicate that a conventional bound state -- 
not a virtual one-- might be found 
for pion masses higher than $290-350$ MeV, 
contrary to the 460 MeV we found at one-loop, which as we have seen was  for sure 
outside the 
region where our approach is reliable. 
This is in qualitative agreement with some recent 
lattice results in \cite{Prelovsek:2010kg},
where they seem to find a bound state 
 for $m_\pi \simeq 325\,$MeV. 
Let us nevertheless recall the caveats raised 
from the very authors of \cite{Prelovsek:2010kg}, since they cannot calculate accurately the width, and some possibly
relevant contributions -- mainly the disconnected contractions-- 
have not been included in the calculation. 
Other lattice studies \cite{Mathur:2006bs} have also suggested 
the existence of a ``tetraquark'' component  for $m_\pi \simeq 180-300\,$MeV.
Let us note, however, that the binding energy of the states 
we find at two-loops seems to grow faster with the pion mass 
than for the one-loop case and on the lattice. However, this occurs already
in the region $m_\pi>350\,$MeV, where we do not consider our approach reliable and the uncertainties are huge as seen by the spread of the fits in Fig.\ref{sigmafigpredictions}. Unfortunately, other relevant lattice calculations for the
$\sigma$ \cite{Alford:2000mm,Kunihiro:2008dc}, lie beyond our reach.

Finally, in the lower panel of Fig.\ref{sigmafigpredictions} we show our results for the $\sigma-2\pi$ coupling,
obtained from the residue of the second Riemann sheet pole -- or poles when there are two branches.
The qualitative behavior is similar to the one-loop case shown in 
Fig.\ref{fig:couplings}, with a dramatic rise up to a peak 
that occurs at the pion mass where the two conjugate poles meet on the real axis. From that value onwards we thus have to draw two branches for each fit, and it can be noticed that the coupling for one of these branches reaches zero. This corresponds to the pion mass where the upper branch pole reaches the $2\pi$ threshold. The fact that at threshold the coupling goes to zero is
in good agreement with the well known 
result in \cite{Weinberg:1962hj}. Actually,
this can be checked analytically because, as shown
in \cite{Gamermann:2009uq}, 
the coupling is inversely proportional to the energy 
derivative of the one-loop function 
which is divergent at threshold.

\section{Summary}
\label{sec:summary}

Using the IAM, which is based on analyticity, elastic unitarity and ChPT,
we generate the poles associated to the $\rho$ and $\sigma$ resonances
without any assumption on their existence or nature. The IAM implements
the pion mass dependence of observables through the subtraction constants 
 up to a given order in ChPT.
Thus, we can predict the dependence of the $\rho$ and $\sigma$ pole positions
on $m_\pi$, as done in \cite{chiralexIAM}. Here we present new results that
were missed in the previous paper.  

First, using the one-loop formalism, we have made a comparison 
of our previous results with some recent lattice data, showing that they are compatible.
We have also calculated the $m_\pi$ dependence of the
 $\rho\pi\pi$ and $\sigma\pi\pi$ effective 
couplings, calculated from the pole residues, finding that the
$\rho\pi\pi$ coupling is almost $m_\pi$ independent, whereas the $\sigma\pi\pi$
coupling shows a strong $m_\pi$ dependence. 

Finally, we have extended to two-loops the modified 
Inverse Amplitude Method formalism to account properly for Adler zeros,
which has been applied then to the  $O(p^6)$ 
calculation.  Although no robust predictions can be made for the $\rho$ mass, 
mostly due to the large uncertainties in the low energy constants,
we have been able to describe the elastic 
scattering phase shift data and lattice results on $f_\pi$ and $a_{20}$
with several fits with fairly reasonable values for such low energy constants, 
and the correct $\bar qq$ leading $1/N_c$ behavior of the $\rho$.

With these fits we have obtained relatively robust predictions for other $\rho$ observables and all $\sigma$ parameters, at least up to $m_\pi\simeq 300-350\,$MeV. In particular, we have confirmed the relatively weak dependence of the $\rho-2\pi$ coupling and the approximate validity of the KSRF relation. Concerning the sigma, whose results are much more robust 
than for the $\rho$ since it has a much weaker dependence on the ChPT 
low energy constants, we have confirmed the appearance of two 
virtual poles for sufficiently high pion masses. 
One of these poles 
becomes a virtual state for $m_\pi$ between, roughly, 300 and 350 MeV.
We hope these results could be of use as a guideline for future extrapolations of lattice results down to physical quark mass values.

\section*{Acknowledgments}
We thank C. Hanhart for useful discussions.
Work partially supported by Spanish Ministerio de 
Educaci\'on y Ciencia research contracts: FPA2007-29115-E,
FPA2008-00592 and FIS2006-03438, 
U.Complutense/Banco Santander grant PR34/07-15875-BSCH and
UCM-BSCH GR58/08 910309. We acknowledge the support 
of the European Community-Research Infrastructure
Integrating Activity
“Study of Strongly Interacting Matter” 
(acronym HadronPhysics2, Grant Agreement
n. 227431)
under the Seventh Framework Programme of EU.

\end{document}